\documentclass[aps,floatfix,onecolumn,prl,superscriptaddress]{revtex4-2}
\usepackage{graphicx,amsmath,amssymb,bbm,xcolor}
\usepackage{multirow,booktabs}
\usepackage{xurl} 

\graphicspath{{main_text_figures/}}

\begin{document}
\title{The Spatial Complexity of Optical Computing and How to Reduce It}

\author{Yandong Li}
 \email{yl2695@cornell.edu}
 \affiliation{School of Electrical and Computer Engineering, Cornell University, Ithaca, New York 14853, USA}
\author{Francesco Monticone}
 \email{francesco.monticone@cornell.edu}
 \affiliation{School of Electrical and Computer Engineering, Cornell University, Ithaca, New York 14853, USA}

\begin{abstract}
Similar to algorithms, which consume time and memory to run, hardware requires resources to function.
For devices processing physical waves, implementing operations needs sufficient ``space," as dictated by wave physics.
How much space is needed to perform a certain function is a fundamental question in optics, with recent research addressing it for given mathematical operations, but not for more general computing tasks, e.g., classification.
Inspired by computational complexity theory, we study the ``spatial complexity" of optical computing systems in terms of scaling laws---specifically, how their physical dimensions must scale as the dimension of the mathematical operation increases---and propose a new paradigm for designing optical computing systems: space-efficient neuromorphic optics, based on structural sparsity constraints and neural pruning methods motivated by wave physics (notably, the concept of ``overlapping nonlocality'').
On two mainstream platforms, free-space optics and on-chip integrated photonics, our methods demonstrate substantial size reductions (to $1\%-10\%$ the size of conventional designs) with minimal compromise on performance.
Our theoretical and computational results reveal a trend of diminishing returns on accuracy as structure dimensions increase, providing a new perspective for interpreting and approaching the ultimate limits of optical computing---a balanced trade-off between device size and accuracy.
\end{abstract}

\maketitle

The introduction of the Turing machine~\cite{Turing:1936} marks a historic tipping point when humankind began to develop a deep mathematical understanding of computational machinery.
Since then, when evaluating whether one problem is harder than another, we can rigorously examine the problems by analyzing their ``computational complexity."
Understanding computational complexity has played an unrivaled role in modern computer science: The pursuit of solving (or approximating) as many problems as possible within polynomially constrained time and memory space not only promotes more efficient implementations but also deepens our understanding of the mathematical and physical mechanisms underlying the algorithms. In this Golden Age of AI, although computational resources have been increasing at an accelerating pace~\cite{ai_and_compute:2018} and deep learning algorithms often prioritize performance over a transparently interpretable formalism~\cite{DL_visual_interpretability:2018,LLM_interpretability:2024}, studying computational complexity is still vital, as these investigations may reveal the scaling law(s) behind  foundation models such as large language models (LLMs), thereby enabling more strategic resource allocation~\cite{scaling_law_OpenAI:2020,scaling_law_DeepMind:2022} and shedding light on their remarkable performance~\cite{emergent_abilities:2022,ZeyuanZhu:2024}.

Over the past decades, the analysis of complexity has been extended to the studies of quantum systems~\cite{linear_optics_complexity:2010,Ising_formulations:2014}, nervous systems~\cite{complex_brain_networks:2009}, and VLSI architecture~\cite{VLSI_design:2011}, aiming to explore new science and optimize engineering solutions.
On the other hand, sensing, imaging, and pre-processing hardware, which are often based on optical (or, more generally, wave-based) platforms, are usually overlooked in complexity analysis because they do not directly perform full-fledged computational tasks.
Nevertheless, the increasing demand for low latency, high bandwidth/throughput, and energy-efficient data processing has promoted developments in optical pre-processing architectures for various applications, including autonomous driving (particularly with LiDARs~\cite{photonic_LiDAR:2022}), augmented and virtual reality (AR/VR)~\cite{AR_glass:2024}, and high-speed biological microscopy~\cite{high_speed_microscopy:2023}.
In each application, light is engineered in a unique, objective-oriented way, and therefore, manipulating light to perform specific functionalities can indeed be considered a form of computation.
Because one of the most critical limiting factors of hardware is its spatial dimensions~\cite{Markov:2014}, one could raise the following general question on the ``computational complexity" of the hardware:
Given a functionality to realize, how small can the optical hardware be?
Or, in other words, what are the fundamental scaling laws that govern the size of a wave-based computing system as a function of the complexity of the task it is designed to perform?

Recently, D. A. B. Miller laid the foundation for these investigations by mapping the desired input-to-output function to a precisely defined concept of ``overlap" between independent ``communication channels'' within the device, which in turn determines size requirements, such as the minimum possible thickness of imaging systems~\cite{Miller:2023}.
These fundamental limits, however, only strictly apply to well-defined, exact mathematical operations (for example, spatial differentiation, i.e., edge detection), whereas it is not clear how these size requirements would apply to more general, less exact computing functions, including inference tasks with tolerable levels of inaccuracy and performance degradation (e.g., classification).
As an example, while it is clear how thin an optical system can be to perform a spatial derivative~\cite{Miller:2023,perspective_Monticone:2023}, it is still unclear how thin it can be to perform image classification.

\begin{figure}[ht]
\centering
    \includegraphics[width=1\textwidth]{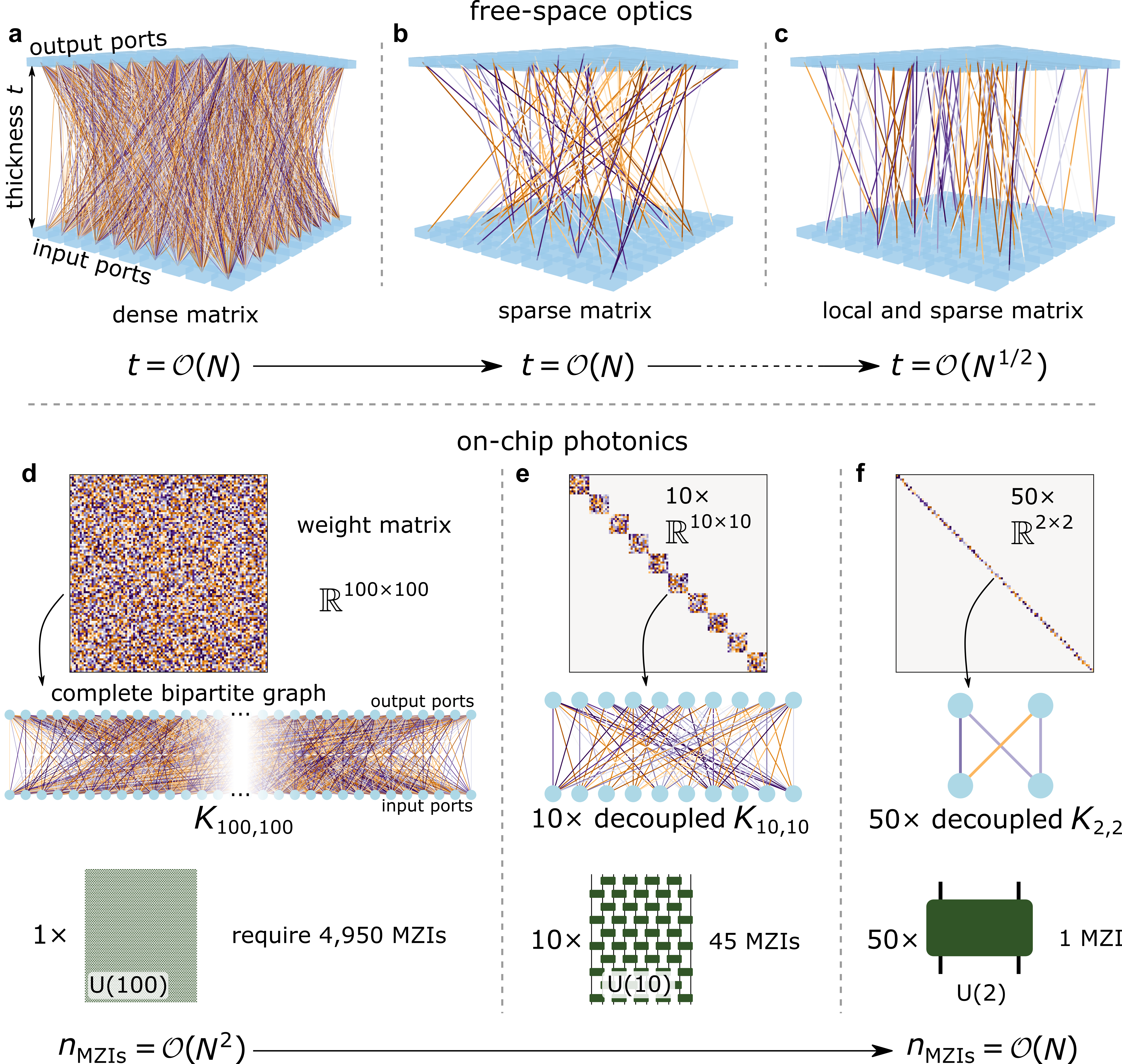}
    \caption{\label{fig:lsmat_and_bloc_diag}
    \textbf{Reducing the spatial complexity of optical computing.}
    Improving the scaling laws of free-space optics (a-c) and photonic chips (d-f) following the proposed space-efficient design paradigm.
    (a-c) Illustration of simplifying a generic free-space optical system.
    The thickness scaling law can be reduced to $\mathcal{O}(N^{1/2})$ when the $N \times N$ kernel matrix, representing the input-output function of the optical system, is designed to exhibit a ``local sparse" form.
    Coloured lines represent the coupling coefficients between input and output ports (sampling points of the corresponding field profiles).
    Visually, a sparse matrix has significantly fewer couplings than a dense one, while in a local kernel matrix, all couplings are only slightly inclined from vertical, whereas nonlocal kernels lack this feature.
    (d-f) Illustration of simplifying a generic two-dimensional photonic chip, composed of a mesh of Mach–Zehnder interferometers (MZIs).
    The block-diagonalization, understood from a graph perspective, breaks the $N \times N$ kernel matrix to a linear number $N/N'$ of small, decoupled complete bipartite graphs, $K_{N',N'}$, each of which requires $N'(N'-1)/2$ MZIs to realize.
    Therefore, when each block is sufficiently small, block-diagonalization reduces the total number of required MZIs from quadratic to quasi-linear.
    }
\end{figure}

To address these fundamental questions and advance one step closer toward the ultimate limits of wave-based computing, we propose and demonstrated the following two-pronged approach:
1. By analyzing the spatial scaling laws of optics, we discover what types of optics can scale optimally, namely, what physical properties and constraints make their spatial dimensions increase slowly as the dimension of the performed mathematical operation increases (Fig.~\ref{fig:lsmat_and_bloc_diag}).
2. Drawing inspiration from modern algorithm design, we propose a new paradigm for optical design, aimed at an optimal balance between performance and spatial resources, i.e., area or volume, through imposing specific constraints that improve the device kernel's scaling.
Two mainstream optical platforms, free-space and on-chip, are considered.

Specifically, we demonstrate that, to reduce the spatial complexity and improve the scaling, the device kernel should exhibit physics- and platform-specific structural sparsity: the ``local sparse" form for three-dimensional free-space optics, which reduces the linear dependence of device thickness on the dimension of the mathematical operation to a square-root dependence, and the ``block-diagonal" form for two-dimensional photonic chips, which reduces the number of required Mach-Zehnder interferometers from quadratic to quasi-linear.

These unique structural sparsity constraints, however, clearly limit the range of mathematical operations that can be realized, as these forms not only reduce the number of parameters but also constrain how the remaining parameters participate in the operation.
Can the target task still be accomplished with high accuracy using only the required forms of operations? This question is beyond the scope of traditional optical design.
Fortunately, artificial neural networks offer a promising solution, as they tend to exhibit considerable redundancy in their parameters~\cite{predicting_params_in_DL:2013}, providing opportunities for pruning---reducing a large number of parameters in a network by selectively removing non-essential neurons and weights, with minimal impact on model accuracy.
Pruning has proven effective across diverse neural network architectures, from the early-stage convolutional neural networks (CNNs)~\cite{optimal_brain_damage:1989} and multilayer perceptrons~\cite{optimal_brain_surgeon:1992} to modern, advanced CNNs~\cite{efficient_NN:2015,efficient_ConvNets:2017} and Transformers, which underpin LLMs~\cite{prune_attention_heads:2019,reduce_transformer_depth:2019,sheared_LLaMA:2024}.

By leveraging the neuromorphic architectures of optical neural networks (ONNs), we translate structural sparsity constraints, originating from wave physics, into optics-specific pruning methods and train local sparse ONNs and block-diagonal ONNs using this approach. We further explore the trade-off between the extent of pruning and the accuracy of these structurally sparse ONNs and identify space-efficient structures that significantly compactify the device (to $1\%-10\%$ the size of conventional designs), with negligible or only slight compromise on accuracy.
Our design paradigm could potentially alleviate part of the computational cost by shifting some operations from electronic to optical processors without a significant increase in device size, thereby paving the way for the development of hybrid optical-electronic edge devices that optimally balance resource consumption and performance.

\section{Results}

\begin{figure}[h]
\centering
    \includegraphics[width=0.8\textwidth]{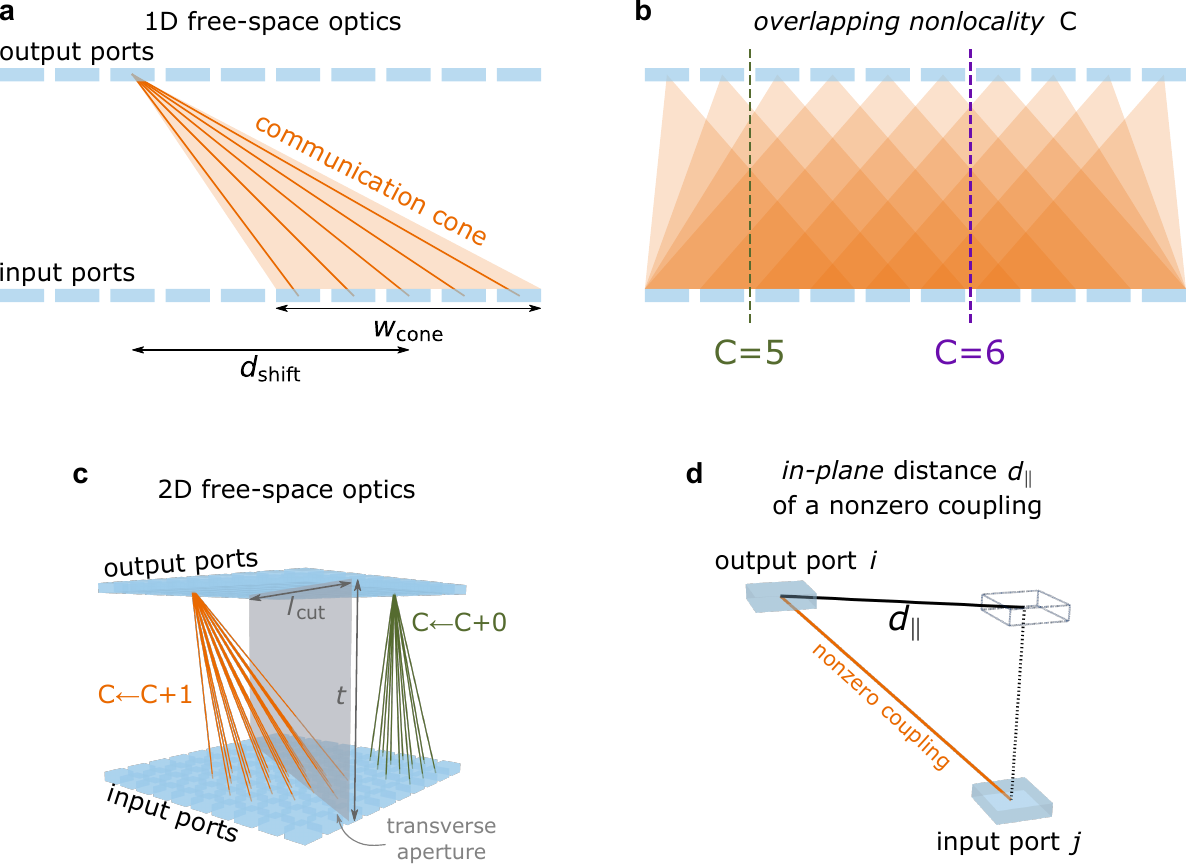}
    \caption{\label{fig:nonlocality}
    \textbf{Standard optical nonlocality and overlapping nonlocality.}
    (a) Schematic of a 1D free-space optical system.
    The \textit{communication cone} of an output port, defined as the set of all couplings to it, is characterized by two parameters, the horizontal shift $d_{\text{shift}}$ and the spanning range $w_{\text{cone}}$.
    An ideally local optical device would have both $d_{\text{shift}}$ and $w_{\text{cone}}$ equal to zero.
    (b) The overlapping nonlocality (ONL) $C$ associated with a transverse aperture (or ``cut") is the number of communication cones intersecting it.
    For example, $C=5$ for the green cut and $C=6$ for the purple cut.
    (c) Demonstration of calculating $C$ for an arbitrary cut for a 2D free-space optical system.
    For a given cut, an output contributes a one to $C$ only if its communication cone intersects the cut.
    (d) Definition of the in-plane distance $d_{\parallel}$ traversed by an optical coupling.
    For a pair of coupled ports $(i, j)$, $d_{\parallel}$ is the distance between them, projected onto the output (or input) plane. This distance measures the nonlocality of each \emph{individual} coupling.
    }
\end{figure}

\begin{figure}[h]
\centering
    \includegraphics[width=1.0\textwidth]{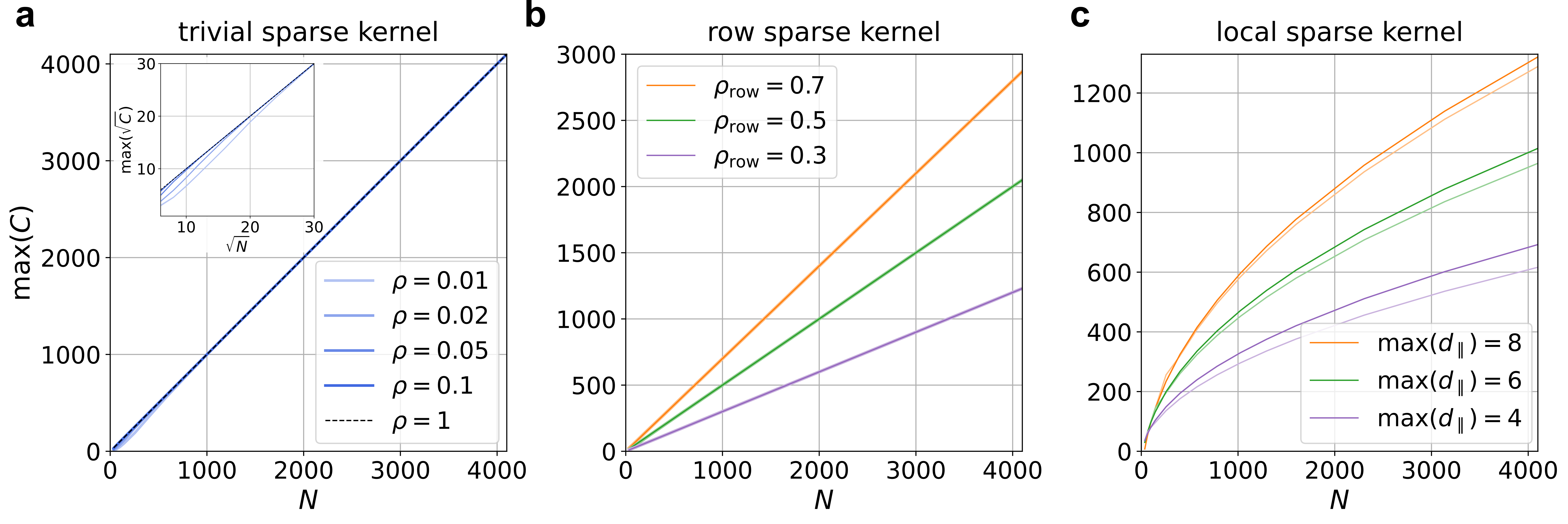}
    \caption{\label{fig:scaling_laws}
    \textbf{Scaling laws of three types of optical device kernels}.
    Scaling laws of the maximum ONL, $\max(C)$, with respect to the mathematical operation dimension, $N$, for the three considered types of device kernels: (a) trivial sparse, (b) row sparse, and (c) local sparse kernels.
    In (a), lines and shaded regions (inset) represent the average and one standard deviation of numerical simulation results.
    The standard deviation is negligibly small.
    In (b,c), darker lines represent theoretical values derived from Eq.~\ref{eq:scaling_laws}, and lighter lines represent numerical simulation results (see Methods and Supplementary Note~1).
    }
\end{figure}

\subsection{The Spatial Complexity of Free-Space Optics}
To implement a nontrivial mathematical operation, an optical device has to be ``nonlocal'' --- its output at any position should depend on the input across a range of positions, which ultimately implies the need for sufficient thickness to incorporate enough ``channels” to communicate sideways within the device~\cite{Miller:2023,review_Monticone:2023}.
The relation between the concept of nonlocality and thickness (and, more broadly, spatial complexity) is, however, a subtle one.

Consider the configuration shown in Fig.~\ref{fig:nonlocality}a: In a 1D optical structure, each output port (a sampling point for the output field) is coupled to a set of input ports (sampling points for the input field), spanning a range over the input plane.
These couplings between pairs of points (input $j$ and output $i$) encode the function of the optical system, namely, they are the elements $D_{ij}$ of its kernel operator (in matrix form) $\mathbf{D}$.
Two parameters can be used to describe how spatially extended, or ``nonlocal'', the range of these couplings is:
$d_{\text{shift}}$, the horizontal shift of the center of this range relative to the output, and $w_{\text{cone}}$, the width of the range.
We then refer to the cone-shaped region (highlighted in orange in Fig.~\ref{fig:nonlocality}a) that encloses all the couplings for an arbitrary output $i$ as the \textit{communication cone} of $i$.
The parameters $d_{\text{shift}}$ and $w_{\text{cone}}$ represent different aspects of the conventional notion of nonlocality in optics and wave physics.
Importantly, however, very large values for these parameters do not imply that the optical system require multiple ``sideways channels” and, thus, a large thickness.
As argued in~\cite{Miller:2023}, a single-mode optical fiber with multiple taps, for example, can have arbitrarily large nonlocality, as quantified here by $d_{\text{shift}}$ and $w_{\text{cone}}$, but would only require a finite (small) thickness to accommodate that single mode.
It is a different aspect of ``nonlocality'' that is directly related to the size requirements of optics, the so-called \textit{overlapping nonlocality} (ONL), which arises if the input position range for one output point overlaps with that for another output point, requiring multiple sideways channels to implement the desired function~\cite{Miller:2023}.

From a geometrical perspective (Fig.~\ref{fig:nonlocality}b), the ONL associated with a transverse aperture $S$ is the number $C$ of communication cones that must cross from one side of $S$ to the other to implement the desired kernel.
We refer to the transverse aperture as a ``cut," as it divides the output (and input) plane into two parts.
For a device operating at an effective wavelength of $\lambda_0/n$ (where $n$ is the maximum refractive index in the device), the cut must be large enough to support $C$ sideways channels to realize those crossing communication cones---one sideways channel for each cone.
Using diffraction arguments (assuming the system only contains transparent, non-absorbing materials), one can then determine the lower bound on the device thickness $t$ from the maximum value of $C$ throughout the device~\cite{Miller:2023}:
\begin{equation}\label{eq:min_thickness}
t \geq 
\begin{cases} 
\max(C) \frac{\lambda_0}{2 (1-\cos\theta) n} & \text{for 1D}, \\
\max(C) \frac{1}{l_{\text{cut}}} \left[ \frac{\lambda_0}{2 (1-\cos\theta) n} \right]^2 & \text{for 2D},
\end{cases}
\end{equation}
where the parameter $\theta$ represents the maximum allowed ray angle inside the device.
Eq.~\ref{eq:min_thickness}, also distinguishes between 1D and 2D cases, since in 1D, the size of the aperture is the thickness $t$, whereas in 2D, $t \equiv A/l_{\text{cut}}$, where $A$ is the area of the cut and $l_{\text{cut}}$ is the length of its projection to the output (or input) plane (Fig.~\ref{fig:nonlocality}c).

These theoretical results imply that the thickness required to perform an exact mathematical operation, described by a kernel $\mathbf{D}$, can only be reduced by decreasing the wavelength or increasing the refractive index, neither of which may be desirable or possible~\cite{limits_on_refractive_index:2011}.
This is valid for any free-space optical (or wave-based) analog computing system, as the ONL $C$ is purely determined by the mathematical form of the desired functionality and not by the details of the optical implementation.

Here, instead, we approach the problem of minimizing the thickness, and therefore the size, of the system from a different perspective---by modifying the form of the mathematical operation so that the thickness scales slowly as the operation dimension increases, while maintaining high accuracy for a given task.
To achieve this, we first analyze the scaling laws of the ONL, $C$, for operations under various constraints, and clarify the relationship between the ONL, sparsity, and standard optical locality.

Consider a device kernel $\mathbf{D}$ of dimension $N \times N$, with input and output ports arranged in grids on two parallel planes (see Supplementary Note~1).
For the same $\mathbf{D}$, the number of required sideways channels, $C$, may vary across different choices of cuts. The most important quantity is therefore $\max(C)$, which is associated with the cut that is crossed by the largest number of sideways channels, and the entire device must be sufficiently thick to accommodate all the associated channels (see Supplementary Note~1 for the definitions of valid and invalid cuts).
To analyze how the form of $\mathbf{D}$ affects $\max(C)$, we first study the worst-case scenario, where every output is coupled to all input points, resulting in a completely filled matrix $\mathbf{D}$. In this case, $C=N$, since the communication cone of every output intersects every cut.

Next, we examine whether increasing sparsity in $\mathbf{D}$ (by introducing zeros, i.e., removing couplings) reduces $\max(C)$.
Our analysis and numerical experiments (see Supplementary Note~1 for the proof and simulation) show that trivial sparsity does not help, except when $\mathbf{D}$ becomes ``extremely sparse," as shown in Fig.~\ref{fig:scaling_laws}a.
However, an overly sparse kernel cannot encode enough free parameters, largely limiting its ability to realize useful optical functions. Therefore, trivially increasing sparsity is not a feasible way to reduce $\max(C)$.

One seemingly promising approach is to impose more drastic structural sparsity by removing a fraction of output ports, as this completely prevents those outputs from contributing to $C$. In matrix terms, removing outputs corresponds to setting rows of $\mathbf{D}$ to zero. With a fraction $\rho_{\text{row}}$ of rows remaining activated,
$\max(C)$ scales as $\rho_{\text{row}} N$, still depending linearly on $N$ (Fig.~\ref{fig:scaling_laws}b).

Aiming for a better scaling law, we then encourage locality when imposing sparsity by constructing matrices whose entries $D_{ij}$ are nonzero only when the \textit{in-plane} distance $d_{\parallel}(i,j)$ (Fig.~\ref{fig:nonlocality}d) is below a predefined threshold $\max(d_{\parallel})$.
This, in turn, constrains all outputs' communication cones, $d_{\text{shift}} + w_{\text{cone}}/2 \leq \max(d_{\parallel})$.
Matrices constructed in this way display a distinct feature:
When viewed in a graph layout, all couplings deviate only slightly from vertical (Fig.~\ref{fig:lsmat_and_bloc_diag}c).
We refer to these matrices as local sparse matrices.
While reducing the in-plane distance crossed by individual couplings does not directly reduce the ONL, if this distance is, on average, small for all couplings, then the number of independent communication cones crossing any transverse aperture should be reduced, hence reducing the ONL.
Our further analysis and simulation indeed reveal that, with local sparse matrices, $\max(C)$ scales sublinearly as $\mathcal{O}(N^{1/2})$ (Fig.~\ref{fig:scaling_laws}c). 
Eq.~\ref{eq:scaling_laws} summarizes the scaling laws for the expected values of the maximum ONL for the three considered types of matrices, corresponding to different forms of device kernel (see Supplementary Note~1 for details and proofs),
\begin{equation}\label{eq:scaling_laws}
\mathbb{E}[\max(C)] = 
\begin{cases}
N
& = \mathcal{O} \left(N\right) \text{ for trivial sparse matrices}, \\
\rho_{\text{row}} N
& = \mathcal{O} \left(N\right) \text{ for row sparse matrices}, \\
2\max(d_{\parallel}) \cdot \left[ \sqrt{2N} - \max(d_{\parallel}) \right]
& = \mathcal{O} \left(N^{1/2}\right) \text{ for local sparse matrices with small $\max(d_{\parallel})$}.
\end{cases}
\end{equation}

\begin{figure}[ht]
\centering
    \includegraphics[width=0.82\textwidth]{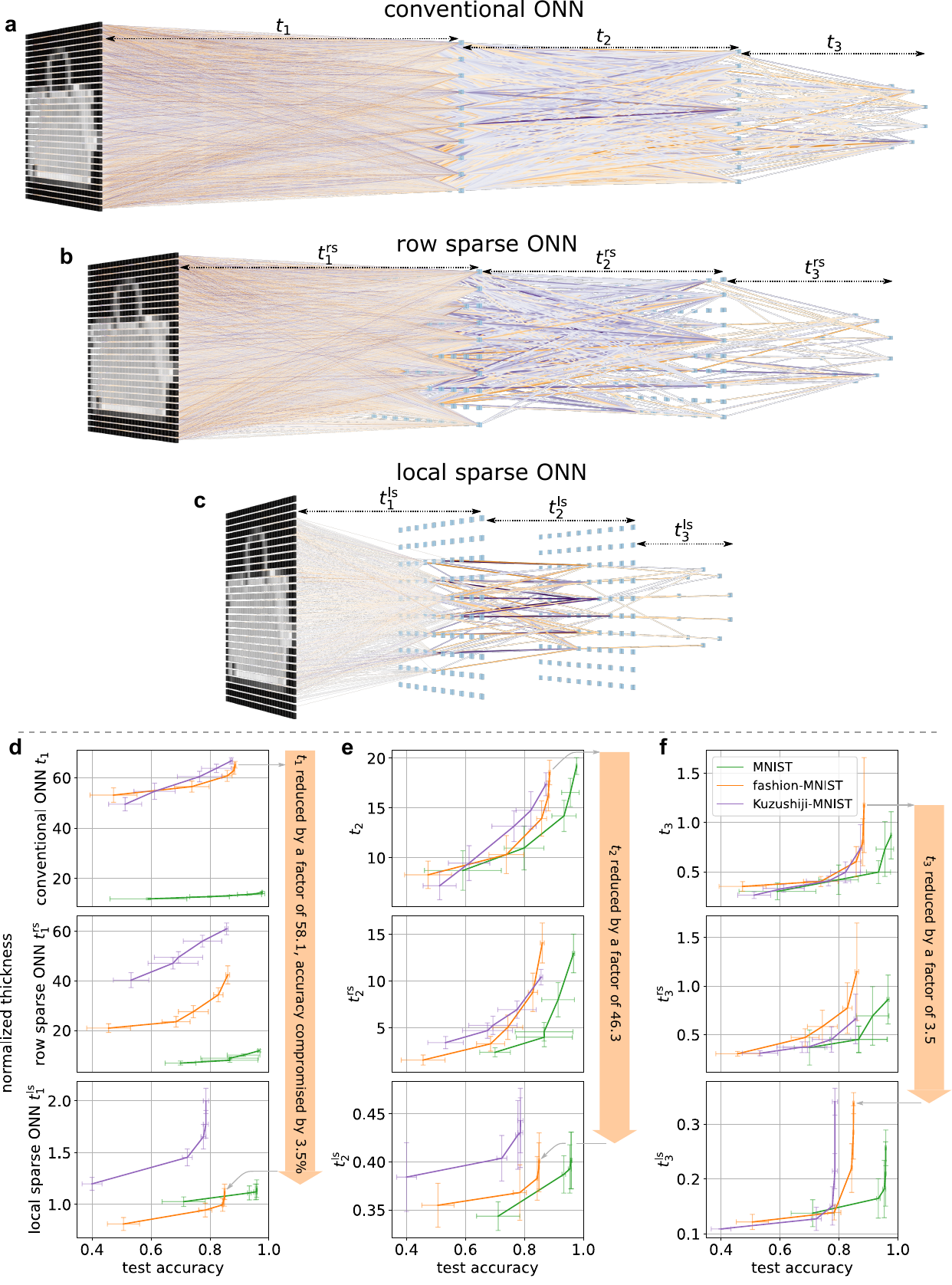}
    \caption{\label{fig:bimt2d_results} \textbf{Space-efficient computing with free-space optics.} Schematic of (a) conventional, (b) row sparse, and (c) local sparse ONNs performing a classification task on the fashion-MNIST dataset.
    (d-f) Thicknesses of the three interlayer regions in conventional (top row), row sparse (middle row), and local sparse (bottom row) ONNs.
    To demonstrate the trade-off between model thickness and accuracy, each model type is pruned using five different pruning thresholds (see Methods).
    Error bars represent one standard deviation of thickness (along $y$-axis) and accuracy (along $x$-axis) across eight different random seeds.
    A local sparse ONN with a pruning threshold of $\tau = 0.01$, compared to a conventional ONN with $\tau = 0.05$, drastically reduces the thickness while only compromising accuracy by $3.6\%$ on the fashion-MNIST dataset (vertical orange arrows;
    see Supplementary Note~5).
    }
\end{figure}

Our findings imply that, while enforcing sparsity alone is not helpful, reducing the degree of standard optical nonlocality, quantified by $d_{\text{shift}}$ and $w_{\text{cone}}$ and bounded by $\max(d_{\parallel})$, also reduces the scaling of the maximum ONL, $\max(C)$, with the operation dimension.
Therefore, reducing the spatial complexity of a free-space optical computing structure requires a pruning method motivated by wave physics and fundamentally rooted in the concept of optical nonlocality.
These results promise compact, scalable optical realizations of large-scale computations, provided that the target task can still be accomplished with high accuracy, as discussed next.

\subsection{Space-Efficient Computing with Free-Space Optics: Local Sparse ONNs}
Unlike artificial neural networks stored and manipulated in computer memory, optical (or, any wave-based) neural networks exist in the physical world.
Many properties of these networks, including their optical locality, stem from the Euclidean distance defined in the position space, thus demanding physics-motivated pruning techniques.
To translate the structural sparsity constraints identified above into an optics-specific pruning method, we leverage the recently proposed brain inspired modular training (BIMT)~\cite{BIMT:2023}, which offers a perfect routine for pruning physical neural networks targeting for local sparse patterns.
BIMT focuses on spatial networks~\cite{review_Barthelemy:2011}, where each neuron in the network is assigned a position, enabling the definition of physical distances. To enhance locality, BIMT penalizes nonlocal channels using an additional loss term, $\mathcal{L}_{\text{nonlocal}}$ (see Methods), and strategically swaps neurons and the associated weights across layers (except for the input).
These features make BIMT an ideal method for designing local sparse optical neural networks (LSONNs), suitable for ultrathin free-space optical devices.

To quantitatively examine the trade-off between the optical device's minimum thickness and the model's inference accuracy, we train LSONNs and their conventional counterparts on three datasets, MNIST~\cite{mnist:2010}, fashion-MNIST~\cite{fashion_mnist:2017}, and Kuzushiji-MNIST~\cite{kuzushiji_mnist:2018}.
Then, we apply weight pruning to the resulting networks by removing weights with absolute values below a threshold $\tau$.
(Here, weight pruning is also applied to conventional networks because, otherwise, the network would be fully connected, corresponding to an extremely thick device.) Each network has $4$ bias-free layers with $\left[ 784, 100, 100, 10 \right]$ neurons. The neurons in the input and hidden layers are arranged in square grids in position space.
The $10$ neurons in the output layer are arranged in a ring.
Each layer, except the last one, is followed by a SiLU (Swish) activation function~\cite{swish_activation:2017}.
Once the network is trained, we calculate the minimum possible device thickness (i.e., the minimum distance between each two successive layers) for a free-space optical implementation, by calculating the maximum ONL and applying Eq.~\ref{eq:min_thickness}.

Fig.~\ref{fig:bimt2d_results} summarizes the thickness-accuracy trade-offs of conventional, row sparse, and local sparse ONNs.
Remarkably, across all three datasets, LSONNs achieve a thickness reduction of more than one order of magnitude in $t_{1,2}$ and reduce $t_3$ by approximately a factor of $3$ compared to their conventional and row sparse counterparts, with a slight or negligible drop in accuracy (e.g., only $3.6\%$ on the fashion-MNIST dataset).
This thickness-accuracy trade-off can be interpreted as diminishing returns on accuracy: beyond a certain level of optical complexity and accuracy, additional spatial resources lead to only marginal improvements in accuracy. Additionally, we attribute this substantial reduction in thickness to a strategic allocation of the ONL across all transverse apertures:
In LSONNs, large values of $C$ always pass through apertures with large in-plane length $l_{\text{cut}}$, while in conventional and row sparse ONNs, large values of $C$ often need to pass through narrow apertures (with small $l_{\text{cut}}$), causing ``information bottlenecks'' that require greater thickness to accommodate the necessary $C$ (see Supplementary Note~2).

In summary, we have demonstrated training LSONNs as an example of the proposed space-efficient paradigm for designing free-space optical computing structures that use an available volume optimally to perform an intended computation task:
First, we have analyzed the scaling laws for the ONL, the key physical quantity associated with the spatial complexity of optical systems, and discovered that mathematical operations in the local sparse form can result in thin free-space optical structures.
Then, by applying a physics-motivated pruning method that targets the required local sparse form, we have trained specialized ONNs that meet the desired criteria.
This procedure is directly applicable to the designs of diffractive-surface- and metalens-based ONNs~\cite{Ozcan:2018,Heide:2023}, potentially enabling a new generation of scalable ultrathin free-space optical devices.

\subsection{Space-Efficient Computing on Photonic Chips: Analysis and Design}
\begin{figure}[ht]
\centering
    \includegraphics[width=1\textwidth]{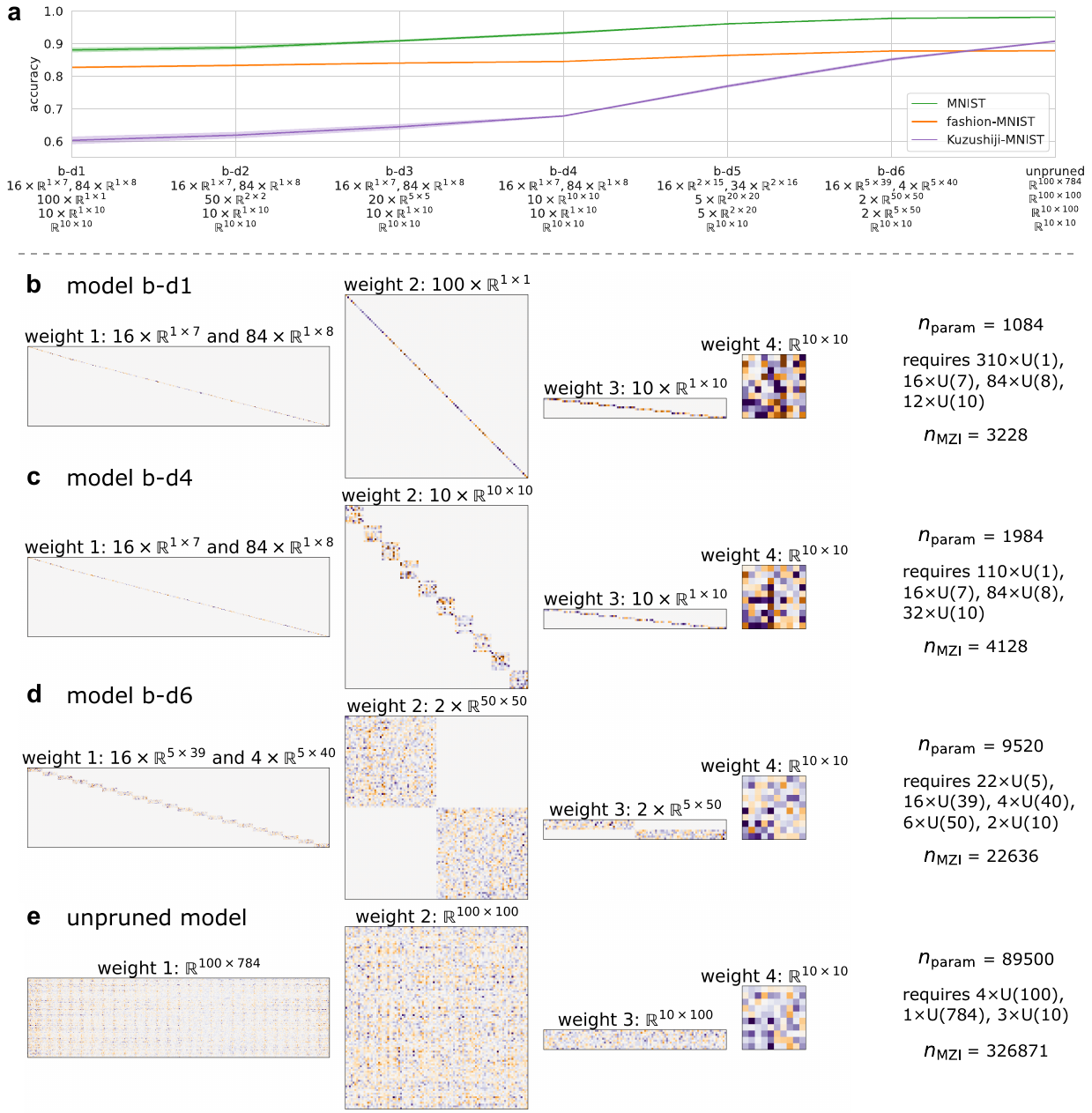}
    \caption{\label{fig:bloc_diag_results}
    \textbf{Space-efficient computing on photonic chips.}
    (a) Trade-off between model accuracy and the degree of block-diagonalization for different block-diagonal models.
    Shaded regions represent one standard deviation of accuracy across eight different random seeds.
    (b-e) Weights of block-diagonal and unpruned models.
    For each model, the number of free parameters, the involved unitary transformations (including both left and right unitary matrices, $U$ and $V^T$, from the singular value decomposition of all weight matrices), and the total number of required MZIs are listed on the right.
    }
\end{figure}

Optical networks on integrated photonic chips are typically based on arrays of photonic waveguides and modulators and meshes of Mach-Zehnder Interferometers (MZIs) performing arbitrary linear operations on the waveguide inputs~\cite{Clements:2016,review_programmable_PC:2020}.
Optimizing spatial resources in this context, therefore, demands reducing the number of required MZIs, $n_{\text{MZI}}$.
When realizing a device kernel matrix $\mathbf{D}$ of dimension $M \times N$, an MZI-based architecture requires $n_{\text{MZI}} = \left[ M(M-1) + N(N-1) \right]/2$ components to implement two unitary transformations, $U(M)$ and $U(N)$~\cite{Clements:2016}.
However, if the kernel matrix could be reduced to a ``block-diagonal" form, the scaling of $n_{\text{MZI}}$ with both $M$ and $N$ would be reduced from quadratic to quasi-linear, as a block-diagonal matrix only comprises a linear number of small blocks with each block only requiring a few MZIs.
This simplification of the photonic chip architecture becomes more apparent from a graph perspective:
Block-diagonalization disassembles the huge bipartite connectivity graph into small, decoupled (disconnected) components, resulting in a significant loss of parameters and, more importantly, inevitably sacrificing inter-block connectivity (coupling), as shown in the connectivity graphs in Fig.~\ref{fig:lsmat_and_bloc_diag}d-f.
Similar to free-space optics, the question is then whether a given operation can still be performed using a space-efficient architecture of this type, and what level of performance degradation can be expected.

While this block-diagonalization approach is not expected to be applicable to all computational tasks, we find that relevant tasks solvable by neural networks are amenable to block-diagonalization, leading to a substantial reduction in spatial complexity.
To optimize the trade-off between $n_{\text{MZI}}$ and the ONNs' performance, we propose a two-step pruning method.
First, we include in the loss function an extra term, $\mathcal{L}_{\text{off-bloc-diag}}$, that penalizes all off-block-diagonal entries.
No parameter is removed at this stage.
Once the weights are trained, they are loaded into specially designed block-diagonal-structured (BDS) linear layers, where only entries in the diagonal blocks are registered as learnable parameters, while all other entries remain zero and are not updatable.
The model with BDS linear layers is further trained on the same dataset to fully adapt to the block diagonal structure (see Methods).

To evaluate the trade-off between $n_{\text{MZI}}$ and the model's inference accuracy, we train networks with several degrees of block-diagonalization on the three MNIST-like datasets used in the previous section~\cite{mnist:2010,fashion_mnist:2017,kuzushiji_mnist:2018}.
Each network has $5$ bias-free layers with $\left[ 784, 100, 100, 10, 10 \right]$ neurons.
Each layer, except the last one, is followed by a SiLU activation function~\cite{swish_activation:2017}.
We consider six different block-diagonalized architectures, named from ``b-d1" to ``b-d6", where $n_{\text{MZI}}$ varies from approximately $3,200$ to approximately $22,600$.
Fig.~\ref{fig:bloc_diag_results} summarizes the trade-off between $n_{\text{MZI}}$ and accuracy.
These six models share one important feature, the weight (size $10 \times 10$) connecting the last two layers remains unpruned, which we find is crucial for maintaining satisfactory accuracy.
(Results on models with a pruned last weight are included in Supplementary Note~5.)
Remarkably, model ``b-d4", with its largest block being only $10 \times 10$ in size, achieves accuracies of $93.24 \pm 0.40\%$, $84.51 \pm 0.16\%$, and $66.76 \pm 0.25\%$ on the MNIST, fashion-MNIST, Kuzushiji-MNIST datasets, respectively.
Implementing ``b-d4" using the MZI-grid architecture already requires fewer than $4,200$ MZIs, compared to the unpruned implementation's approximately $0.3$ million MZIs---a reduction of about $98.7\%$.
We also expect that emergent photonic components, for example, inverse-designed, programmable multiport couplers~\cite{LiangFeng:2023,MoLi:2024,Onodera:2024}, could enable an even more compact and flexibly tunable on-chip implementation of similar models based on small, decoupled blocks.

Finally, we extend this block-diagonalization approach to 1. develop block-circulant ONNs, whose weight matrices consist of rows of cyclically shifting blocks, leveraging light diffraction for space-efficient computing (e.g.,~\cite{space_efficient_optics:2022});
and 2. prune models optimized for edge AI applications, e.g., MobileNetV2~\cite{MobileNetV2:2018} for CIFAR-10~\cite{cifar10:2009} classification.
We block-diagonalize the fully connected (FC) layer of MobileNetV2 and, with a largest block size of $10 \times 10$, reduce $n_{\text{MZI}}$ by $\sim 99\%$ (considering the photonic implementation of FC layers only), while the accuracy decreases only slightly from $92.97 \pm 0.24\%$ to $92.41 \pm 0.25\%$.
(See Supplementary Note~5 for the model architecture.)
With these extensions, we envision that a hybrid architecture---employing digital implementations for convolution layers, nonlinear activations, shortcut connections, etc., and photonic analog components for the classifier based on block-diagonal (or block-circulant) FC layers---could facilitate FLOP-, memory-, and space-efficient computation acceleration in real-world edge applications.

\section{Discussion}
For future optical systems to help addressing the growing challenge of large-scale, ultrafast, energy-efficient parallel data processing in a scalable way, we believe their spatial complexity should be carefully assessed to avoid trading off the benefits afforded by optics with impractical size requirements.
Specifically, we have shown that, to create optical systems that scale optimally with the operation dimension, their kernel operators need to exhibit physics- and platform-specific structurally sparse patterns: the ``local sparse" form for free-space optics and the ``block-diagonal" form for photonic chips. Our space-efficient design paradigm offers engineering solutions aligned with significantly improved scaling laws, as demonstrated by examples of local sparse and block-diagonal ONNs.
The space-accuracy trade-offs in these space-efficient ONNs reveal a trend of diminishing returns on accuracy: once the optics reaches a certain level of complexity and accuracy, further investments in spatial resources yield marginal accuracy increase.
In fact, we have identified optimal ONNs that could achieve comparable accuracy to unpruned networks while consuming only $1\%-10\%$ of their spatial resources.
In light of this, we interpret the ultimate limits of optical computing as a balance between device dimensions and performance, rather than merely high performance metrics.
To achieve an optimal space-accuracy trade-off, adjusting the global connectivity topology of the optical system is as important as fine-tuning well-understood optical degrees of freedom, such as the refractive index contrast---if not more than.

Furthermore, space-efficient ONNs offer the first example of exploiting the full potential of neural pruning with optical hardware.
Although the pruning-induced sparsity offers great flexibility and enables high compression rates for neural networks, leveraging this sparsity to achieve fast inference typically requires specialized software (e.g., cuSPARSE~\cite{cusparse}) or hardware (e.g., EIE~\cite{EIE:2016}) accelerators~\cite{Zhuang_Liu:2017}.
This is because the memory access patterns in sparse operations may not fully utilize GPUs' parallel architecture.
Consequently, while a sparse matrix has fewer effective operations, the overhead of managing sparsity and irregular memory access patterns can offset these benefits.
This mismatch is an example of poor synergy between the model and the computational hardware.
We believe that optical neuromorphic analog computing, with its high energy efficiency, high throughput, and low latency~\cite{review_McMahon:2023,comment_Li_Monticone:2024}, when fully optimized for space efficiency, could be a promising candidate for achieving an excellent software-hardware synergy and ultimately leading to superior computational acceleration.

We also note that the resources used in optical computing are not limited to spatial dimensions.
Systems operating on synthetic dimensions, such as frequency~\cite{Senanian:2023,review_Yuan:2018}, angular momentum~\cite{review_Yuan:2018}, and time~\cite{Marandi:2024}, would follow different spatial scaling laws, but we expect that any reduction in spatial complexity would be accompanied by increased computational complexity in other dimensions (e.g., operation time scaling as $\mathcal{O}(N^2)$) and other trade-offs (e.g., time-accuracy trade-offs).
The study of multi-dimensional scaling laws for wave-based computing systems remains an interesting open area of research.
At last, we note that the two mentioned optical/photonic platforms were neither originally intended for neuromorphic computing nor for emulating complex (brain-like) systems.
As depicted in Fig.~\ref{fig:lsmat_and_bloc_diag}, the connectivity within these architectures is highly regular, resulting in a simple graph topology compared to complex networks, such as power grids, brain networks, or social networks~\cite{Barabasi:1999,review_Strogatz:2001,Bassett:2006}.
These complex networks can exhibit various unique phenomena, such as the small-world effect~\cite{Watts:1998}, where each node can reach any other node through only a few intermediary ``repeater" nodes, even if the network is extremely large, indicating that these networks are highly efficient in information exchange~\cite{efficient_small_world:2001,Jiaxuan_You:2020}.
Could more intricate connectivity with richer topology enable even more space-efficient optics and reduce the spatial complexity scaling law below polynomial growth?
Highly scattering media~\cite{review_Gigan:2017} may be a candidate that could facilitate such complex connectivity in a wave-based system.

\section{Acknowledgements}
This work was supported by the Air Force Office of Scientific Research with grant no. FA9550-22-1-0204 and Office of Naval Research with grant no. N00014-22-1-2486.
Y. Li thanks Prof. Jon Kleinberg for helpful discussions on information networks.

\section{Data and Code Availability}
All datasets used in this research are publicly available.
The code for generating and analyzing sparse, row sparse, and local sparse matrices is available at \url{https://github.com/lyd5039/The-Spatial-Complexity-of-Optical-Computing/}.

\section{Competing Interests}
Y. L. and F. M. are listed as inventors on a U.S. provisional patent application (Serial No.~63/718,499) on training and pruning methods for designing space-efficient free-space optics and photonic chips.

\section{References}
\bibliographystyle{unsrt}

\newpage
\section{Methods}
\subsection{Analyzing the Scaling Laws of the Three Types of Matrices}
The algorithms for generating sparse, row sparse, and local sparse matrices are detailed in Supplementary Note~1, with the corresponding pseudocode provided in Supplementary Note~4.
The numerical method for calculating $\max(C)$ is detailed in Supplementary Note~1.

After proving that introducing trivial (non-structural) sparsity into the kernel matrix does not affect $\max(C)$, we explore row sparse matrices characterized by a single parameter $\rho_{\text{row}}$, the ratio of activated rows.
Although row sparse matrices can be made even sparser by setting some entries in the activated rows to zero (thus involving an additional parameter, the density $\rho$), increasing sparsity in this trivial, non-structural manner does not affect $\max(C)$ for large matrices either (see Supplementary Note~1).
Therefore, for Fig.~\ref{fig:scaling_laws}b, we activate all entries within activated rows in numerical simulations (using the `all possible' setting in \texttt{Row Sparse Matrix Constructor} in Supplementary Note~4).

In local sparse matrices, all matrix entries $M_{ij}$ that satisfy the local condition $d_{\parallel}(i,j) \leq \max(d_{\parallel})$ are activatable.
Similarly, in the numerical simulations for Fig.~\ref{fig:scaling_laws}c, we activate all local entries (using the `all possible' setting in \texttt{2D Local and Sparse Matrix Constructor}).
Notably, trivially increasing sparsity by setting some local entries to zero can further reduce $\max(C)$ (see Supplementary Note~1).

\subsection{BIMT for 2D Local Sparse ONNs}
We train 2D local sparse ONNs using the following loss function~\cite{BIMT:2023},
\begin{equation}\label{eq:total_loss_LSONN}
\mathcal{L} = \mathcal{L}_{\text{MSE}}
+ \lambda_{\text{nl}} \mathcal{L}_{\text{nonlocal}},
\end{equation}
where $\mathcal{L}_{\text{MSE}}$ is the mean squared error loss for prediction, and
\begin{equation}\label{eq:bimt_loss}
\mathcal{L}_{\text{nonlocal}} \equiv \sum^L_{l=1} \sum^{N^{(l)}}_{i=1} \sum^{N^{(l-1)}}_{j=1} d^{(l)}_{\parallel}(i,j) |D^{(l)}_{ij}|
\end{equation}
which sums the product of the modulus of the weight $D^{(l)}_{ij}$ associated with every coupling and the ``in-plane" distance $d^{(l)}_{\parallel}(i,j)$ it crosses, over all layers $l \in \{ 1,2,\cdots,L \}$ except the input layer ($l=0$).
$N^{(l)}$ represents the number of neurons in layer $l$.
We set all layers to be bias-free, so $\mathcal{L}_{\text{nonlocal}}$ does not contain the bias term.
We train the model with $\lambda_{\text{nl}} = 0.02$ initially and schedule it to increase during training.

\subsection{Pruning the Three Types of ONNs}
In Fig.~\ref{fig:bimt2d_results}d-f, we have demonstrated the trade-off between model thickness and accuracy using the following five pruning settings applied to conventional, row sparse, and local sparse ONNs:
\begin{itemize}
    \item \textbf{Conventional ONNs} are pruned with five different threshold values: $\tau = 0.05, 0.075, 0.1, 0.15, \text{ and }0.2$.
    Entries in the weight matrices are set to zero if their absolute values are below $\tau$.
    \item \textbf{Row sparse ONNs} are constructed by pruning unimportant neurons from the two hidden layers.
    \textit{Neuron importance} is defined as the activation output $\mathbf{a}^{(l)}$ of the forward pass of layer $l$ (where $l=1,2$ correspond to the two hidden layers),
    \begin{equation}
    \mathbf{a}^{(l)} = \sigma(\mathbf{D}^{(l)} \mathbf{a}^{(l-1)} + \mathbf{b}^{(l)}).
    \end{equation}
    Here, $\mathbf{b} = \mathbf{0}$ since the bias terms are disabled, and the nonlinear activation function $\sigma(\cdot)$ is the SiLU function.
    If the absolute value of the $j$-th element, $|a_j^{(l)}|$, falls below a predefined threshold $\tau_\text{neuron}$, the corresponding neuron is considered unimportant and pruned.
    This is done by setting the $j$-th row of this layer's weight (optical device kernel) $\mathbf{D}^{(l)}$ to zero, ensuring that the $j$-th neuron's output remains zero in subsequent forward passes after pruning.
    We calculate the neuron importance on a randomly selected $1/5$ subset ($10,000$ samples) of the training set and compute the average neuron importance over this subset.
    
    The five thresholds for unimportant neurons are $\tau_\text{neuron} = 0.05, 0.075, 0.1, 0.125, \text{ and }0.15$.
    After removing unimportant neurons, we apply weight pruning as before:
    weight matrix entries with absolute values below $\tau = 0.05, 0.075, 0.1, 0.1, \text{ and }0.15$ are set to zero.
    In summary, neurons with importance below $\tau_\text{neuron}$ are pruned first, followed by pruning weights below $\tau$, resulting in row sparse ONNs.
    
    \item \textbf{LSONNs} are pruned using threshold values $\tau = 0.01, 0.02, 0.05, 0.1, \text{ and }0.2$.
    Notably, even with weight pruning alone, a significant percentage of neurons in the two hidden layers are deactivated, making additional neuron pruning unnecessary for LSONNs (see Supplementary Note~1).
\end{itemize}

For the same model and dataset, higher values of thresholds $\tau$ and $\tau_{\text{neuron}}$ correspond to more aggressive pruning, leading to thinner interlayer regions.
In each subfigure in Fig.~\ref{fig:bimt2d_results}d-f, the data points with lower thickness correspond to higher pruning thresholds.

\subsection{Calculating the Physical Thicknesses of ONNs}
To calculate the thicknesses of the pruned conventional, row sparse, and local sparse ONNs, we first normalize the distance between adjacent neurons, $d_{\text{adj}}$, so that each layer fits within a square of area $28^2 = 784$.
Thus, in the input layer, the distance between adjacent ports is normalized to the unit length, $d_{\text{adj}}^{(l=0)} = 1$.
For the two hidden layers, the adjacent distance is set to $d_{\text{adj}}^{(l=1,2)} = 2.8$, and the output layer is resized similarly.
Then, we sweep over all \textit{valid} cuts (see Supplementary Note~1) between each pair of adjacent layers and identify the cut associated with the $\max(C/l_{\text{cut}})$.
This approach ensures a consistent comparison since $\max(C)$ alone is proportional to the cut area, $A_{\text{cut}}$, where $A_{\text{cut}} = l_{\text{cut}} t$~\cite{Miller:2023}.

\subsection{Two-Step Pruning for Block-Diagonal NNs}
We train block-diagonal NNs in the following two phases.

\textbf{Phase I}: the model is trained using the following loss function,
\begin{equation}\label{eq:total_loss_obd}
\mathcal{L} = \mathcal{L}_{\text{CE}}
+ \lambda_{\text{obd}} \mathcal{L}_{\text{off-bloc-diag}},
\end{equation}
where $\mathcal{L}_{\text{CE}}$ is the cross entropy loss for prediction, and
\begin{equation}\label{eq:obd_loss}
\mathcal{L}_{\text{off-bloc-diag}} \equiv \sum^L_{l=1} \sum_{(i,j) \not\in \text{diag blocks}} |D^{(l)}_{ij}|
\end{equation}
for the weight $\mathbf{D}^{(l)}$ of all $l \in \{ 1,2,\cdots,L \}$ layers.

\textbf{Phase II}: The weights learned during Phase I are loaded into block-diagonal-structured (BDS) models, where only the diagonal blocks contain learnable parameters.
We then train the BDS models, during which the off-block-diagonal entries consistently remain zero.
Phase II serves as an optional fine-tuning process.

The block-diagonal models ``b-d1" to ``b-d5" are trained through the two phases.
Model ``b-d6" is processed only the first phase, as progressing to the second phase results in a slight reduction in accuracy.
This decrease is likely attributable to overfitting, given that ``b-d6" has a relatively high number of parameters.

\subsection{Pruning MobileNetV2}
Given that the original MobileNetV2~\cite{MobileNetV2:2018} is designed for real-world image datasets like ImageNet (resolution $\sim 256\times256$), we employ a slightly simplified version of MobileNetV2~\cite{MobileNetV2_cifar10} for classifying the CIFAR-10 dataset~\cite{cifar10:2009}.
Details of the architecture are provided in Supplementary Note~5.

In the unpruned model, the output (of dimension $1 \times 1 \times 1280$) from the last layer (avgpool $8 \times 8$) is reshaped and fed into a FC layer of dimension $10 \times 1280$ for CIFAR-10 classification.
With a single layer, the dimension is reduced so drastically that block-diagonalization does not help much in reducing the number of required MZIs.
Even block-diagonalizing this layer to $10 \times \mathbb{R}^{1 \times 128}$, each of the ten unitary matrices $U(128)$ would still require more than $8,000$ MZIs to implement.
To address this problem, in the pruned model, we deepen the FC layers to five layers with $[1250, 250, 50, 10, 10]$ neurons (the output dimension of the avgpool layer is changed to $1250$).
With the addition of three layers, the dimension reduction is done smoothly.
The layer weights are block-diagonalized to $125 \times \mathbb{R}^{2 \times 10}$, $25 \times \mathbb{R}^{2 \times 10}$, $5 \times \mathbb{R}^{2 \times 10}$, and $\mathbb{R}^{10 \times 10}$, thus ensuring that the largest unitary matrix involved is $U(10)$, significantly reducing the spatial complexity.

\end{document}


\title{Supplementary Material for\\ ``The Spatial Complexity of Optical Computing and How to Reduce It''}

\author{Yandong Li}
 \email{yl2695@cornell.edu}
 \affiliation{School of Electrical and Computer Engineering, Cornell University, Ithaca, New York 14853, USA}
\author{Francesco Monticone}
 \email{francesco.monticone@cornell.edu}
 \affiliation{School of Electrical and Computer Engineering, Cornell University, Ithaca, New York 14853, USA}

\maketitle
\tableofcontents
\section{Scaling Laws of Free-Space Optics}
In this section, we discuss the scaling laws, specifically how the maximum \textit{overlapping nonlocality}, denoted as $\max(C)$, scales with the matrix dimension $N$ for trivial sparse, row sparse, and local sparse matrices. To provide a clear context for this analysis, we first explain the construction of these matrices in our numerical experiments and describe the numerical recipe for sweeping over different transverse apertures (``cuts" that divide the structure into two).
we present the theoretical derivations of the scaling laws.

\subsection{Definitions of the Three Types of Matrices}
\begin{figure}[ht]
\centering
    \includegraphics[width=0.7\textwidth]{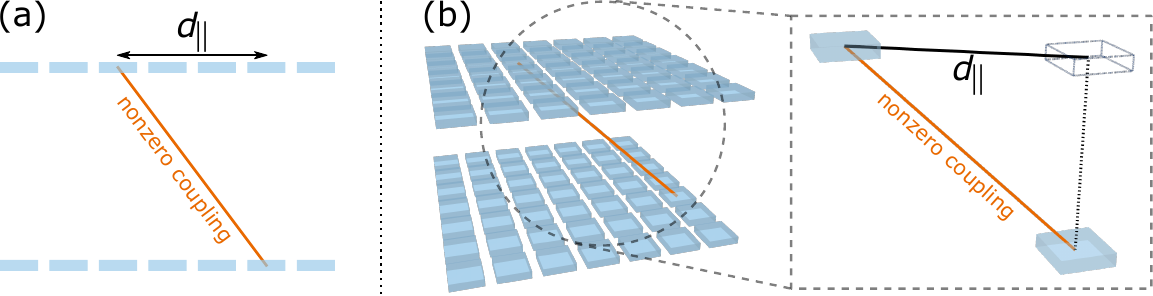}
    \caption{\label{fig:define_d_parallel__1D_2D}
    Definitions of the \textit{in-plane} distance, $d_{\parallel}$, for a nonzero coupling in (a) 1D and (b) 2D configurations.
    }
\end{figure}

Two principal characteristics of these matrices—sparsity and locality—are defined as follows.
Consider matrices of dimension $N_{\text{out}} \times N_{\text{in}}$:
\begin{itemize}
  \item \textbf{Sparsity}: The density of a matrix $M$ is defined as
\begin{equation}
\rho(M) = \frac{\text{number of nonzero elements in } M}{\text{total number of elements in } M}
\end{equation}
Sparse matrices have low $\rho$ values.
  \item \textbf{Locality}: As a geometrical property, locality arises from the spatial arrangement of input and output ports.
  We embed the input (output) ports in a 2D square grid, with a separation of 1 unit between adjacent ports (unless an alternative embedding is specified), as this is the most straightforward way to discretize the input (output) plane.
  For example, the positions of output port $i$ and input port $j$ are
  \begin{equation}\label{eq:2d_grid}
  \begin{aligned}
  \text{input $j$:} \quad & \mathbf{r}_j \equiv (x_j, y_j) \in \left([1, n_{\text{in},x}] \times [1, n_{\text{in},y}]\right) \cap \mathbb{Z}^2, \\
  \text{output $i$:} \quad & \mathbf{r}_i \equiv (x_i, y_i) \in \left([1, n_{\text{out},x}] \times [1, n_{\text{out},y}]\right) \cap \mathbb{Z}^2,
  \end{aligned}
  \end{equation}
  where $n_{\text{in},x} \cdot n_{\text{in},y} = N_{\text{in}}$ and $n_{\text{out},x} \cdot n_{\text{out},y} = N_{\text{out}}$.

  After assigning a spatial coordinate to each input and output, we can calculate the \textit{in-plane} distance, $d_{\parallel}$ (Fig.~\ref{fig:define_d_parallel__1D_2D}), between any pair of input and output ports.
  \textit{Local} matrices satisfy that, 
  for all pairs of input $j$ and output $i$, for an edge to exist, i.e., $|M_{ij}| > 0$, $d_{\parallel}(i,j)$ must not exceed a predefined maximum value:
  \begin{equation}
  \begin{gathered}
  \forall i, j : \ |M_{ij}| > 0 \implies d_{\parallel}(i,j) \leq \max(d_{\parallel}), \\
  \text{where } d_{\parallel}(i,j) \equiv |\mathbf{r}_j - \mathbf{r}_i| = \sqrt{(x_j-x_i)^2+(y_j-y_i)^2}.
  \end{gathered}
  \end{equation}
\end{itemize}

With these definitions in place, we can now describe the construction of the three types of matrices.
\begin{itemize}
    \item \textbf{Trivial sparse matrix}: A specified number of entries are randomly selected across the entire matrix and set to one, ensuring a predefined density $\rho$ of nonzero entries.
    \item \textbf{Row sparse matrix}: We first activate $(\rho_{\text{row}} \cdot N_{\text{out}})$ rows, then randomly populate entries within these rows with ones, achieving the desired density $\rho$ of nonzero entries.
    \item \textbf{Local sparse matrix}: For each entry in the matrix, we calculate its in-plane distance, $d_{\parallel}$. A random subset of entries satisfying $d_{\parallel} \leq \max(d_{\parallel})$ is set to one, achieving the desired density $\rho$ of nonzero entries.
\end{itemize}

These construction methods are detailed in the following algorithms (Algorithms \ref{alg:TrivialSparseMat}, \ref{alg:RowSparseMat}, and \ref{alg:2DLSmat}).
For row sparse matrices and local sparse matrices, the argument $\rho$ can be set to `all possible,' meaning all entries that satisfy the respective conditions are populated.
Without loss of generality, all matrices constructed for analyzing the scaling law are square matrices ($N_{\text{in}} = N_{\text{out}} = N$).

\subsection{Numerical Method for Calculating max(\textit{C})}
With the matrices constructed, we proceed to the numerical experiments of calculating the overlapping nonlocality, $C$, by sweeping over all transverse apertures (valid ``cuts") and counting how many output ports have \textit{communication cones} that cross the cut.
Here, the communication cone of an output port refers to all the couplings that connect this port to the input plane.
If any coupling crosses the cut, the communication cone is considered to cross the cut.
Additionally, it is important to note that only valid cuts are associated with transverse apertures, ensuring the accurate calculation of $C$.

\begin{figure}[ht]
\centering
    \includegraphics[width=0.5\textwidth]{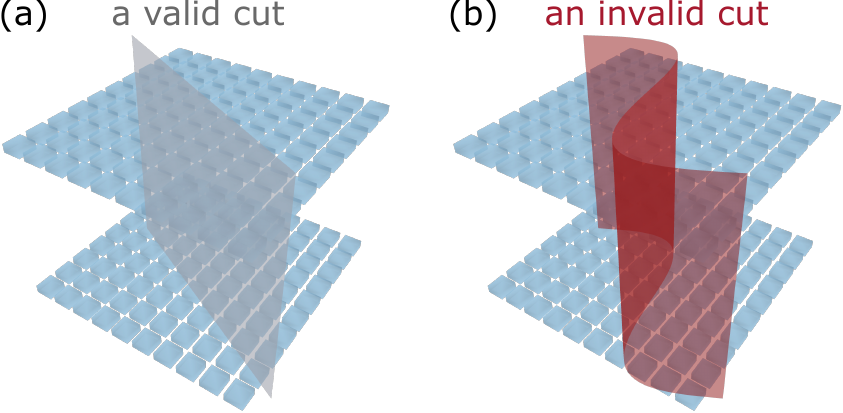}
    \caption{\label{fig:valid_and_invalid_cuts}
    Examples of (a) a valid cut and (b) an invalid cut.
    }
\end{figure}

\textbf{Define and sweep over valid cuts}. Valid cuts are perpendicular to the plane of input ports (and to the plane of output ports).
It is straightforward to find all valid cuts for 1D layouts: when inputs and outputs are equal in number ($N_{\text{in}} = N_{\text{out}} = N$), there are only $N-1$ valid cuts.
However, for 2D layouts, there are many ways to make vertical cuts.
Among all vertical cuts, those whose projection onto the input (or output) plane is not a straight line are considered invalid (Fig.~\ref{fig:valid_and_invalid_cuts}). This is because a communication cone can cross such a cut multiple times, causing overcounting.
Additionally, a valid cut must partition the structure into two non-trivial subsets, ensuring that neither subset contains all ports and couplings while the other remains empty.

\begin{figure}[ht]
\centering
    \includegraphics[width=0.25\textwidth]{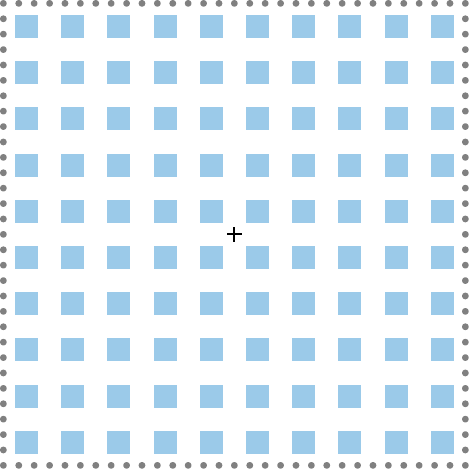}
    \caption{\label{fig:mesh_periphery}
    Mesh the periphery to $2N_{\text{per port}} (n_x + n_y)$ points, where $n_x$ and $n_y$ are the number of ports along $x$ and $y$.
    Here $n_x = n_y = 10$, $N_{\text{per port}} = 3$.
    }
\end{figure}

\begin{figure}[ht]
\centering
    \includegraphics[width=0.99\textwidth]{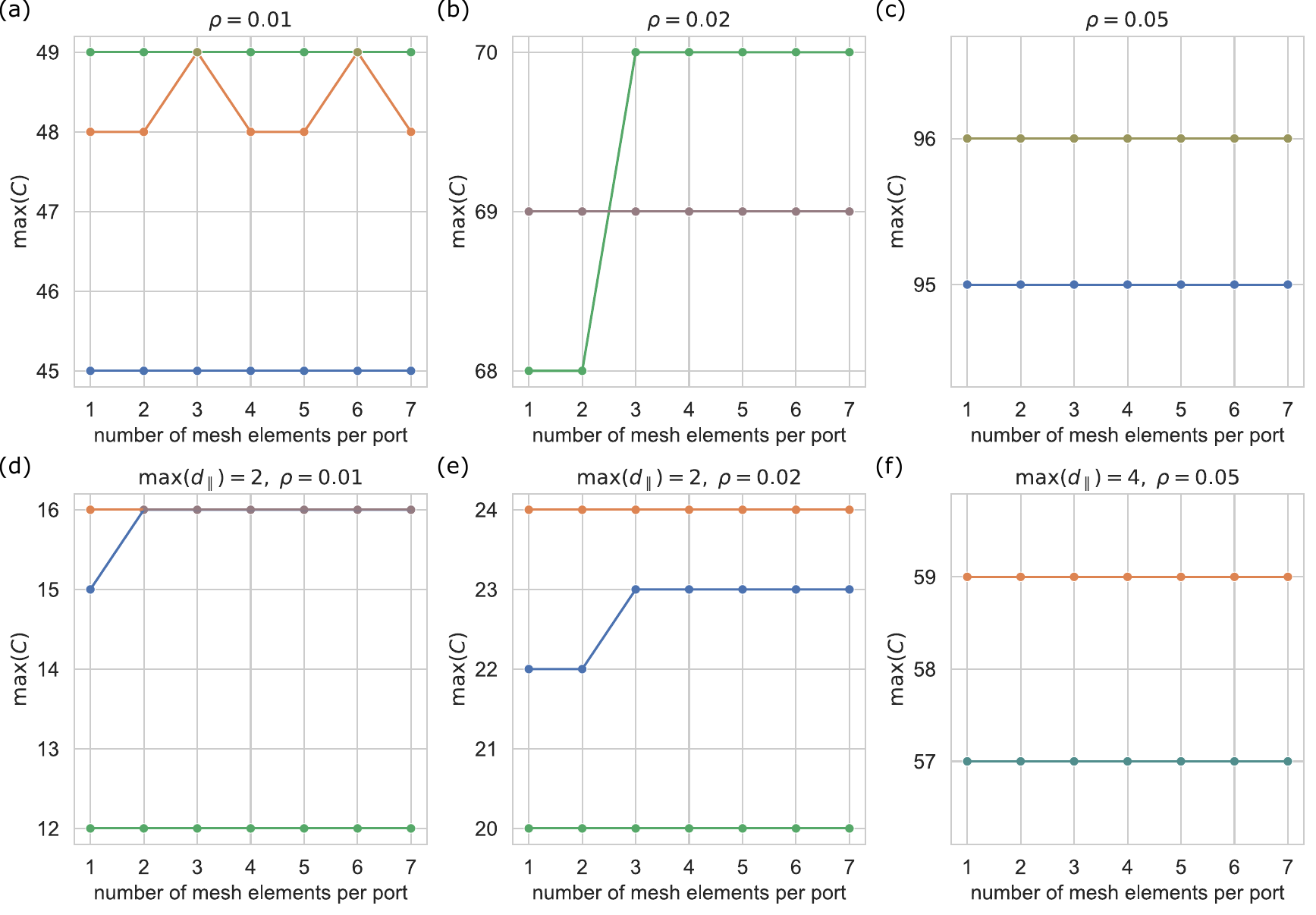}
    \caption{\label{fig:sweep_over_mesh_densities}
    The calculated maximum overlapping nonlocality, $\max(C)$, varies with different mesh densities (represented as the number of mesh elements per port, $N_{\text{mesh per port}}$) around the square periphery.
    (a-c) Randomly generated sparse matrices without locality constraints.
    (d-f) Randomly generated local sparse matrices.
    The three colors represent three different random seeds. Mixed colors represent overlapping lines.
    }
\end{figure}

To sweep all valid cuts for 2D layouts, we mesh the periphery of the entire square layout into $2N_{\text{per port}} (n_x + n_y)$ points and sweep over all vertical cuts that pass through any two mesh points (Fig.~\ref{fig:mesh_periphery}).
Here, $n_x = n_y = \sqrt{N_{\text{in}}} = \sqrt{N_{\text{out}}} = \sqrt{N}$.
(If $N_{\text{in}} \neq N_{\text{out}}$, we use the $n_x, n_y$ such that $n_x n_y = \max(N_{\text{in}}, N_{\text{out}})$.)
Clearly, the number of all vertical cuts is of the order $\mathcal{O}(N_{\text{per port}}^2 \cdot N)$.
We find that setting $N_{\text{per port}} = 3$ is the optimal choice, as a finer mesh does not alter the calculated result of the maximum overlapping nonlocality, $\max(C)$, while a sparser mesh might fail to detect some cuts with high $C$ values (Fig.~\ref{fig:sweep_over_mesh_densities}).

Due to the large time consumption of sweeping over all the $\mathcal{O}(N_{\text{per port}}^2 \cdot N)$ valid cuts for large matrices, we only sweep over \textit{balanced} cuts~\cite{balanced_cut} when numerically experimenting on the constructed matrices to find $\max(C)$.
Balanced cuts pass through the center point of the 2D grid layout (the ``+" mark in Fig.~\ref{fig:mesh_periphery}).
This reduces the number of sweeps to $\mathcal{O}(N_{\text{per port}} \cdot \sqrt{N})$.
We apply this simplification only when numerically analyzing the scaling laws.
When numerically calculating $\max(C)$ and thicknesses of the trained optical neural networks, we sweep over all valid cuts.

\subsection{Theoretical Analysis of the Scaling Laws}
Finally, we delve into the theoretical analysis of the trivial sparse, row sparse, and local sparse matrices and derive the scaling laws.
\begin{equation}\label{eq:scaling_laws}
\mathbb{E}[\max(C)] = 
\begin{cases}
N
& = \mathcal{O} \left(N\right) \text{ for trivial sparse matrices}, \\
\rho_{\text{row}} N
& = \mathcal{O} \left(N\right) \text{ for row sparse matrices}, \\
2\max(d_{\parallel}) \cdot \left[ \sqrt{2N} - \max(d_{\parallel}) \right]
& = \mathcal{O} \left(N^{1/2}\right) \text{ for local sparse matrices with small $\max(d_{\parallel}$)}.
\end{cases}
\end{equation}

\textbf{Trivial sparse matrices}.
For a completely dense matrix (i.e., $\rho = 1$), the overlapping nonlocality, $C$, reaches its maximum possible value, $N$, the output dimension (or more precisely, the smaller of the input and output dimensions~\cite{Miller:2023}), as every output's communication cone crosses every cut.
We prove that, for a kernel matrix exhibiting trivial sparsity---without any specific patterns of zeros---$C$ remains equal to $N$, so the overlapping nonlocality is identical to that of the completely dense matrix, provided the matrix is not ``extremely sparse."

In optical terms, unless the kernel matrix is ``extremely sparse," the free-space optics implementing it will be as thick as those used to realize a completely dense matrix.
We will quantitatively define the threshold at which a matrix is considered ``extremely sparse."

\subsubsection{Introducing trivial sparsity in kernel matrix cannot lead to thinner optics}
Consider a matrix of dimension $N \times N$ with a density $\rho$.
The probability that an arbitrary matrix entry $M_{ij}$ is nonzero is
\begin{equation}
P(M_{ij} \neq 0) = \rho, \quad \text{for all pairs } (i, j).
\end{equation}

\begin{figure}[ht]
\centering
    \includegraphics[width=0.65\textwidth]{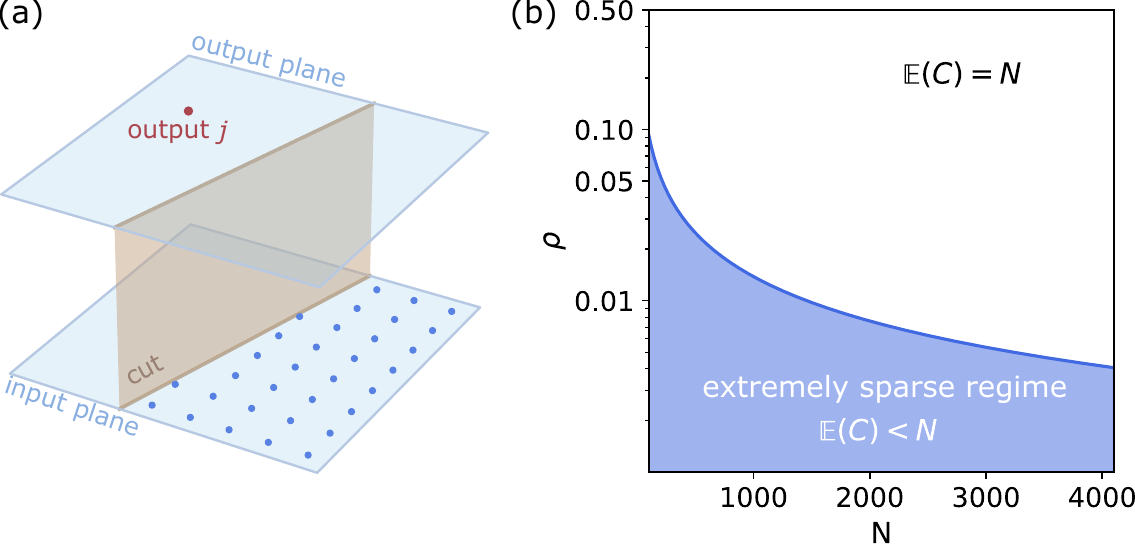}
    \caption{\label{fig:trivial_sparse_mat_property}
    (a) Schematic of a balanced cut dividing the input and output planes.
    For an arbitrary output port, $N/2$ input ports are located on the other side of the cut.
    (b) The ``extremely sparse" regime for matrices with different output dimensions $N$.
    }
\end{figure}

We analyze, for an arbitrary output port $i$, whether its communication cone contributes a one to the overlapping nonlocality $C$ of an arbitrary balanced cut, i.e., whether it crosses the cut.
With such a cut, $N/2$ input ports are not on the same side as the output port $i$ (Fig.~\ref{fig:trivial_sparse_mat_property}a).

Therefore, the number of couplings between the output $i$ and the input plane and cross the cut follows the Binomial distribution, $\text{Binomial} (N/2, \rho)$.
For the output $i$ to contribute a one to $C$, it must have at least one coupling that crosses the cut.
The probability that $i$ has at least one such coupling is
\begin{equation}
P(\text{an output has at least one coupling that crosses the cut}) = 1 - (1-\rho)^{N/2}.
\end{equation}
This can be approximated as
\begin{equation}
P(\text{an output has at least one coupling that crosses the cut})
= 1-(1-\rho)^{N/2}
\approx 1 - e^{-\rho N/2}.
\end{equation}
Given there are $N$ output ports, the expected number of $C$ is
\begin{equation}
\mathbb{E}(C) = N P(\text{an output has at least one coupling that crosses the cut})
\approx N \left(1 - e^{-\rho N/2}\right).
\end{equation}

Now we analyze the condition $\mathbb{E}(C) < N$, i.e., the overlapping nonlocality is at least one less than the output dimension.
This condition is equivalent to $Ne^{-\rho N/2} > 1$, leading to
\begin{equation}
\rho < \frac{2\ln(N)}{N}.
\end{equation}
This defines the ``extremely sparse" condition (Fig.~\ref{fig:trivial_sparse_mat_property}b).
To illustrate how sparse matrices must be to satisfy this criterion, we provide two examples: For $N = 1000$, $\rho$ must be less than $0.014$ for $\mathbb{E}(C)$ to fall below $N$;
for $N = 4000$, $\rho$ must be less than $0.005$.
When $\rho$ is higher than this extremely sparse threshold, $\mathbb{E}(C)$ equals $N$, and consequently, $\mathbb{E}[\max(C)] = N$ since $C$ cannot exceed the number of output ports, $N$.

Therefore, reducing the optical thickness through trivially enhancing the matrix sparsity proves impractical.
As the matrix dimension $N$ increases, even a reduction of one in the overlapping nonlocality $C$ demands an extremely sparse kernel matrix.
When a matrix contains too many zeros, it becomes nearly impossible to embed a useful operation within it.
This extreme sparsity limits the range of mathematical operations that the optical element can effectively implement, thereby significantly restricting its practical application.

\textbf{Row sparse matrices}.
Since trivially increasing sparsity---without imposing any structured patterns of zeros---does not effectively reduce $\max(C)$, we shift our focus to strategically allocating zeros within the matrix.

For a given cut, preventing an output from contributing to $C$ requires that it does not couple with any input located on the opposite side of the cut.
However, as we sweep over different cuts, the inputs on the opposite side change.
A straightforward approach to reducing $C$ is to disable all couplings to a particular output, effectively removing that output.
In matrix terms, this means setting an entire row to zeros, which leads us to the study of row sparse matrices.

In a row sparse matrix with $\rho_{\text{row}} N$ activated rows and a total of $\rho N^2 \leq \rho_{\text{row}} N^2$ nonzero entries, an analysis similar to the previous discussion applies.
The number of couplings between an arbitrary output port and the input plane, crossing a balanced cut, still follows a Binomial distribution, $\text{Binomial} (N/2, \rho/\rho_{\text{row}})$.
Here, the number of trials remains $N/2$ because the number of input ports on the opposite side of the cut is unchanged, as the column dimension is not reduced.
However, the probability is rescaled to $\rho/\rho_{\text{row}}$ since only entries in activated rows can be nonzero:
Given that the total number of nonzero elements in $M$ is $\rho N^2$, the probability that an activatable matrix entry $M_{ij}$ is nonzero is
\begin{equation}
P(M_{ij} \neq 0) = \frac{\rho N^2}{\rho_{\text{row}} N \cdot N} = \frac{\rho}{\rho_{\text{row}}}, \quad \text{for pairs } (i, j) \text{ s.t. } i \in \{\text{activated rows}\}.
\end{equation}

The number of activatable outputs is $\rho_{\text{row}} \cdot N$, so the expected value of $C$ is
\begin{equation} \mathbb{E}(C) \approx \rho_{\text{row}} N \left[ 1 - e^{-\rho N/(2\rho_{\text{row}})} \right]. \end{equation}

When $\rho/\rho_{\text{row}}$ exceeds the extremely sparse threshold, $\mathbb{E}(C)$ approaches $\rho_{\text{row}}$, and therefore $\mathbb{E}[\max(C)] = \rho_{\text{row}} N$, since $C$ cannot exceed the number of activated output ports.
In summary, $\mathbb{E}[\max(C)]$ for row sparse matrices still scales linearly with $N$, but with an additional multiplier $\rho_{\text{row}}$.
This analysis holds as long as $\rho_{\text{row}}$ is not too low (e.g., $\rho_{\text{row}} \geq 0.1$).
In practice, zeroing out too many rows in the kernel matrix may cause problems, as $\text{rank}(M) \leq \rho_{\text{row}} N$, and finding a low-rank approximation of the desired matrix operation may be challenging.

\textbf{Local sparse matrices}.
Both trivial sparse and row sparse matrices are non-geometrical, in that they are defined purely based on matrix density without considering spatial arrangement.
These matrices can be constructed without the need to arrange input and output ports in space.
However, a transverse aperture (``cut") is inherently spatial, as is the associated overlapping nonlocality $C$.
To reduce $C$ from a geometrical or spatial perspective, we propose local sparse matrices (LSmats), where a matrix entry $M_{ij}$ can only be nonzero if the in-plane distance $d_{\parallel}(i,j)$ between the coupled input $j$ and output $i$ is less than a predefined threshold, $\max(d_{\parallel})$.

\begin{figure}[t]
\centering
    \includegraphics[width=0.99\textwidth]{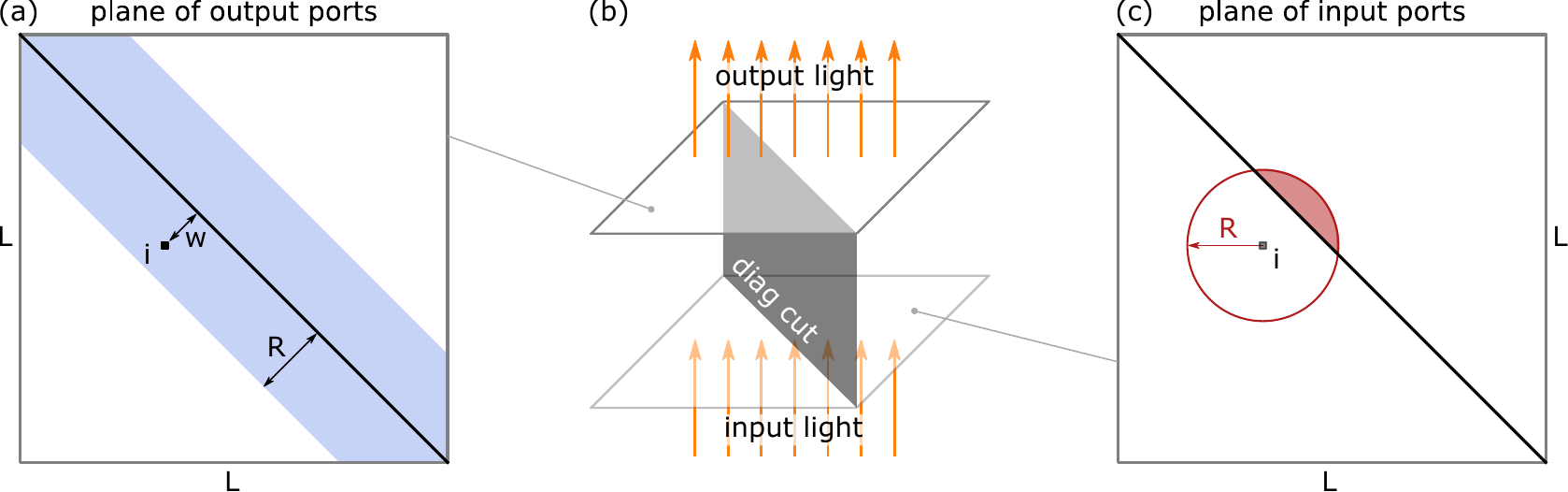}
    \caption{\label{fig:schematic_diag_cut__LSmat}
    Consider a transverse aperture along the diagonal of a free-space optical computing setup.
    Input and output ports are arranged on two planes in a 2D square grid layout. 
    (a) On the output plane, the output port $i$ is located $w$ units from the diagonal cut.
    Only output ports within a distance $w < R \equiv \max(d_{\parallel})$ (shaded light blue) can have communication cones that cross the diagonal cut.
    (b) The diagonal cut divides the input (output) plane into two parts.
    (c) All input ports to which the output $i$ can possibly couple are within the red circle of radius $R$, centered at the projected position of $i$ onto the input plane.
    Only coupling with at least one input port in the red-shaded circular segment can cause $i$ to contribute a one to $C$.
    The side length of the input (output) plane is $L \equiv \sqrt{N-1} \approx \sqrt{N}$.
    }
\end{figure}

The advantage of LSmats becomes apparent in the following example: a cut that divides the input and output planes along the diagonal.
Only output ports close enough to the cut can contribute a one to $C$ (Fig.~\ref{fig:schematic_diag_cut__LSmat}a, blue shaded region).
Consider an arbitrary output port $i$.
On the input plane, all possible ports that can couple to the output $i$ are within the red circle.
Only those in the red-shaded region can cause the output $i$'s communication cone to cross the cut.
However, calculating the exact probability that output $i$ couples with at least one input in the red-shaded region makes the analysis too complicated to proceed with.

Here, we analyze the LSmats whose \textit{all} local entries (i.e., those where $d_{\parallel} \leq \max(d_{\parallel})$) are set to one.
For this diagonal cut, $C$ equals the number of output ports in the blue-shaded region, as each of these ports has its communication cone crossing the cut.
This $C$ is $\max(C)$ because, given $R \equiv \max(d_{\parallel})$, the diagonal cut maximizes the blue-shaded region.
In summary,
\begin{equation}\label{eq:scaling_law_LSmats}
\mathbb{E}[\max(C)] = 2 \left( \sqrt{2}L-R \right) R
= 2\max(d_{\parallel}) \cdot \left[ \sqrt{2N} - \max(d_{\parallel}) \right].
\end{equation}

Eq.~\ref{eq:scaling_law_LSmats} is valid for $0 \leq \max(d_{\parallel}) \leq \sqrt{N/2}$.
When $\max(d_{\parallel})$ reaches $\sqrt{N/2}$, the blue-shaded region (Fig.~\ref{fig:schematic_diag_cut__LSmat}a) spans the entire output plane, meaning every output contributes a one to the $C$ of the diagonal cut.
In fact, when $\max(d_{\parallel}) = \sqrt{N}/2$, $\mathbb{E}[\max(C)]$ already reaches $N$.
Consider a cut that bisects the output plane from one side to the opposite, both passing through the midpoint.
For all outputs, the furthest distance to this cut is $\sqrt{N}/2$, thus $\max(d_{\parallel}) = \sqrt{N}/2$ is sufficient for $\mathbb{E}[\max(C)] = N$.

\begin{figure}[t]
\centering
    \includegraphics[width=0.7\textwidth]{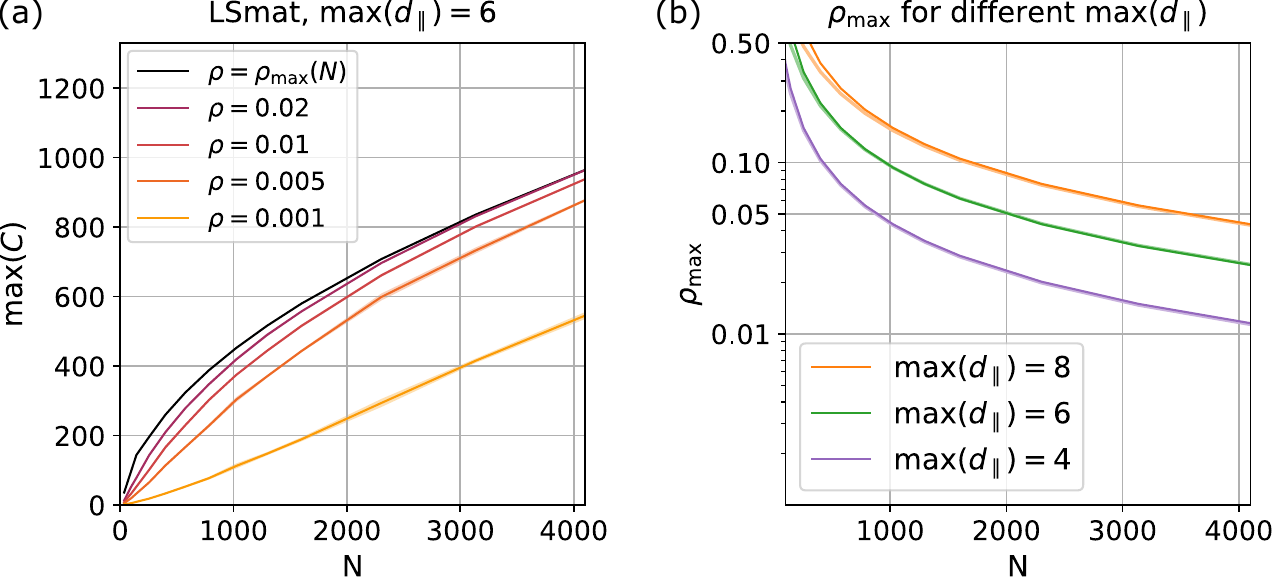}
    \caption{\label{fig:local_sparse_mat_property}
    (a) The maximum overlapping nonlocality $\max(C)$ scales with $N$ for LSmats with different densities.
    Lines and shaded areas represent the average and one standard deviation of numerical simulation results, calculated across eight different random seeds.
    $\rho_{\max}(N)$ denotes the maximum possible density of an $N \times N$ LSmat (with all local entries set to $1$).
    $\max(d_{\parallel})$ is fixed at $6$ throughout these numerical experiments.
    (b) $\rho_{\max}(N)$ for LSmats with different values of $\max(d_{\parallel})$.
    Light-colored lines represent numerical results;
    Dark-colored lines represent theoretical calculations as detailed in Eq.~\ref{eq:max_rho__LSmat}.
    }
\end{figure}

Eq.~\ref{eq:scaling_law_LSmats} shows that LSmats successfully reduce the scaling of $\mathbb{E}[\max(C)]$ below linear dependence on $N$ even when all the local entries are nonzero.
If only a fraction of these entries are nonzero, numerical experiments demonstrate that $\max(C)$ can be further reduced (Fig.~\ref{fig:local_sparse_mat_property}a).

Notably, the density $\rho$ of LSmats is relatively low.
We now delve into this property and address the key question:

\subsubsection{How many free parameters can be encoded in a local sparse matrix?}
Only \textit{local} entries in LSmats are activatable, but how many are there?
If there are not enough nonzero entries, LSmats face the same problem as trivial sparse matrices in the ``extremely sparse" regime: a sufficient number of free parameters cannot be encoded in the matrix, significantly limiting its practical application.

\begin{figure}[t]
\centering
    \includegraphics[width=0.99\textwidth]{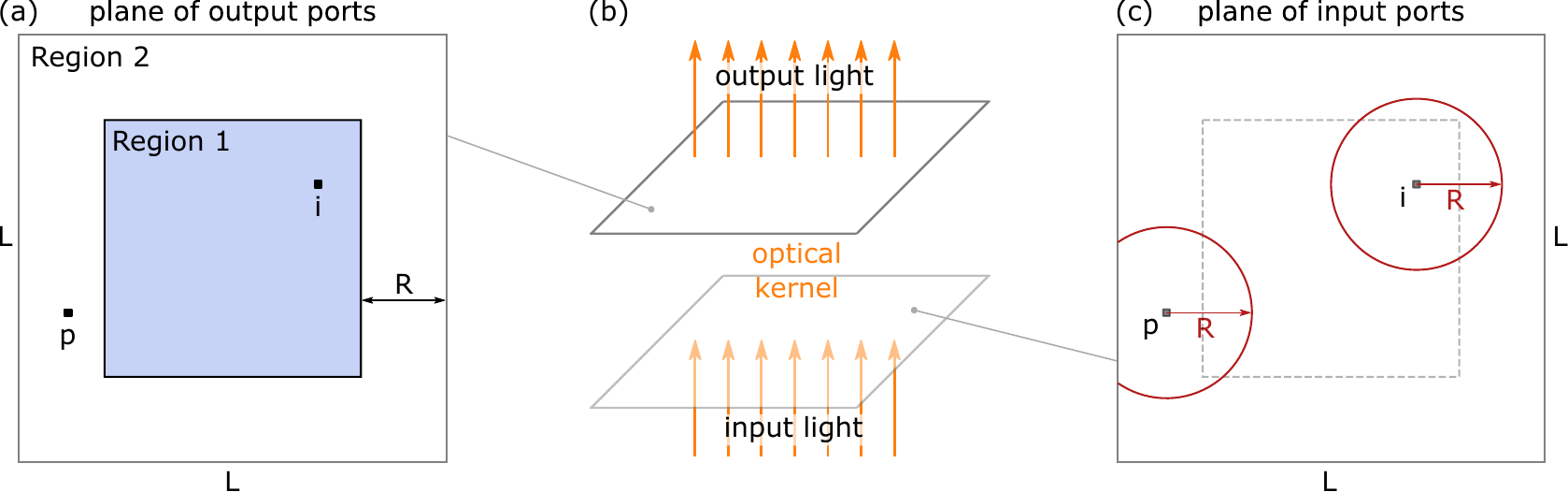}
    \caption{\label{fig:schematic_calculate_max_rho__LSmat}
    (a) The output plane is categorized into two regions:
    Region 1: The communication cone of each output fully overlaps with the input plane.
    Region 2: Only part of the communication cone of each output overlaps with the input plane.
    (b) Schematic of the free-space optical computing setup.
    (c) On the input plane, the communication cones of two representative output ports, output $i$ in Region 1 and output $p$ in Region 2.
    }
\end{figure}

The maximum number of free parameters (represented as $\rho_{\max} \cdot N^2$) in an LSmat should scale linearly with $N$ and quadratically with $\max(d_{\parallel})$, since the number of outputs is $N$ and each can couple to $\pi \max(d_{\parallel})^2$ inputs.
(However, the exact value of $\rho_{\max} N^2$ is slightly lower due to boundary effects.)

To more precisely calculate $\rho_{\max}(N)$ for LSmats of different dimensions $N \times N$, we note that there are two types of output ports (in Regions 1 and 2 in Fig.~\ref{fig:schematic_calculate_max_rho__LSmat}a).
The communication cone of each output in Region 1 completely falls onto the input plane.
Each such cone is associated with $\pi R^2 = \pi [\max(d_{\parallel})]^2$ activatable matrix entries.
However, for outputs in Region 2, part of their communication cones extends beyond the input plane (Fig.~\ref{fig:schematic_calculate_max_rho__LSmat}c).
We approximate these cones as being scaled down by a factor of $3/4$:
Each output in Region 2 is associated with $3 \pi [\max(d_{\parallel})]^2 /4$ activatable matrix entries.
Hence, the maximum possible density $\rho_{\max}(N)$ is
\begin{equation}\label{eq:max_rho__LSmat}
\begin{aligned}
\rho_{\max}(N)
& = \frac{1}{L^4} \left[ (L-2R)^2 + \frac{3}{4}\left( L^2 - (L-2R)^2 \right) \right] \pi R^2
= \left( L^2 - LR + R^2 \right) \frac{\pi R^2}{L^4} \\
& = \left(\frac{1}{N} - \frac{\max(d_{\parallel})}{N^{3/2}} + \frac{[\max(d_{\parallel})]^2}{N^2} \right) \cdot \pi [\max(d_{\parallel})]^2
\end{aligned}
\end{equation}

Thus, we reach an accurate estimate of the maximum number of free parameters in an LSmat.
Since the dominating term of $\rho_{\max}$ scales quadratically with $\max(d_{\parallel})$ (which represents nonlocality), by strategically choosing a small $\max(d_{\parallel})$, we can allocate a sufficient number of free parameters in LSmats.

\section{Why can LSONN achieve a significant reduction in thickness?}
\begin{figure}[ht]
\centering
    \includegraphics[width=0.4\textwidth]{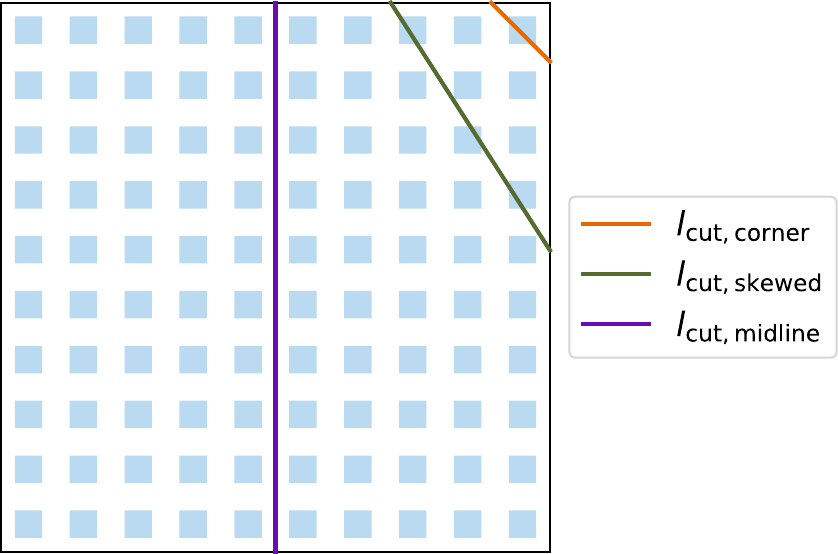}
    \caption{\label{fig:draw_3_characteristic_cuts}
    Examples of a corner cut, a skewed cut, and a midline cut in a 100-input, 100-output optical system, viewed from the vertical direction.
    The input (output) ports are arranged in a $10 \times 10$ grid.
    }
\end{figure}

We observe that local sparse optical neural networks (LSONNs) optimally allocate the overlapping nonlocality $C$ among all transverse apertures (``cuts").
Specifically, to accommodate all the nonlocality in space, the device must be thicker than $\max(t)$, where the maximum is taken over all possible valid cuts.
However, the required thickness $t \propto C / l_{\text{cut}}$ becomes large when the transverse aperture length $l_{\text{cut}}$ is small.
For the trained conventional and row-sparse ONNs, we find that $\max(t)$ (more precisely, $\max(C/l_{\text{cut}})$) is attained at a corner.
That is to say, these two types of ONNs allocate $C$ in an injudicious way, causing an ``information bottleneck" at the corner:
a large amount of $C$ needs to cover the one input port at the corner, whereas the transverse aperture separating that input port from the rest of the system is very narrow, resulting in a large required thickness.
In contrast, for LSONNs, $\max(t)$ is attained at a midline cut for $t_2$ and $t_3$, and at a skewed cut for $t_1$, thus avoiding the bottleneck problem at the corners.
Therefore, the required thickness for LSONN is low because $C$ is allocated prudently, such that a large $C$ corresponds to an aperture with large $l_{\text{cut}}$ within the network.

\section{ONNs with Block-Circulant Weights}
In this section, we revisit a unique property of conventional circulant matrices: they can be diagonalized by the discrete Fourier transform (DFT) matrices.
We then extend this concept to block-circulant matrices, in which each row consists of cyclically shifting blocks, and explore their use as neural network weights for a novel pruning approach.

A circulant matrix can be realized optically using a linear number of MZIs and two diffractive cells that perform Fourier transform operations.
If the circulant matrix $M_{\text{circ}}$ (of dimension $N \times N$) is specified by the vector $\mathbf{c} = [c_0, c_1, \dots, c_{N-1}]$, then $M_{\text{circ}}$ is given by
\begin{equation}
M_{\text{circ}} = \begin{pmatrix}
c_0 & c_{N-1} & c_{N-2} & \cdots & c_2 & c_1 \\
c_1 & c_0 & c_{N-1} & \cdots & c_3 & c_2 \\
c_2 & c_1 & c_0 & \cdots & c_4 & c_3 \\
\vdots & \vdots & \vdots & \ddots & \vdots & \vdots \\
c_{N-2} & c_{N-3} & c_{N-4} & \cdots & c_0 & c_{N-1} \\
c_{N-1} & c_{N-2} & c_{N-3} & \cdots & c_1 & c_0
\end{pmatrix}
\end{equation}

Circulant matrices can be diagonalized by the DFT matrix $F$.
The eigendecomposition of $M_{\text{circ}}$ is given by
\begin{equation}
M_{\text{circ}} = F_N \Lambda F_N^\dagger,
\end{equation}
where $F_N$ is the $N \times N$ DFT matrix, defined as:
\begin{equation}
F_N = \frac{1}{\sqrt{N}} \begin{pmatrix}
1 & 1 & 1 & \cdots & 1 \\
1 & \omega & \omega^2 & \cdots & \omega^{N-1} \\
1 & \omega^2 & \omega^4 & \cdots & \omega^{2(N-1)} \\
\vdots & \vdots & \vdots & \ddots & \vdots \\
1 & \omega^{N-1} & \omega^{2(N-1)} & \cdots & \omega^{(N-1)(N-1)}
\end{pmatrix}
\end{equation}
with $\omega = e^{-2\pi i / N}$.
$\Lambda$ is a diagonal matrix containing the eigenvalues of $M_{\text{circ}}$,
\begin{equation}
\Lambda = \text{diag}(\sqrt{N} F_N^{\dagger} \mathbf{c}).
\end{equation}

In the optical implementation, $F_N$ and $F_N^{\dagger}$ are realized using the diffractive cells, while $\Lambda$ is realized using $N$ MZIs~\cite{space_efficient_optics:2022}.

\textbf{Extension to block-circulant matrices}.
Consider a block-circulant matrix $M_{\text{bloc-circ}}$ composed of $b$ blocks, where each block $B_i$ is a matrix of size $p \times q$.
The block-circulant matrix (of dimension $bp \times bq$) is defined as
\begin{equation}\label{eq:def_bloc_circ_mat}
M_{\text{bloc-circ}} =
\begin{pmatrix}
B_0 & B_{b-1} & B_{b-2} & \cdots & B_2 & B_1 \\
B_1 & B_0 & B_{b-1} & \cdots & B_3 & B_2 \\
B_2 & B_1 & B_0 & \cdots & B_4 & B_3 \\
\vdots & \vdots & \vdots & \ddots & \vdots & \vdots \\
B_{b-2} & B_{b-3} & B_{b-4} & \cdots & B_0 & B_{b-1} \\
B_{b-1} & B_{b-2} & B_{b-3} & \cdots & B_1 & B_0
\end{pmatrix}
\end{equation}

$M_{\text{bloc-circ}}$ can be block-diagonalized by two unitary matrices~\cite{bloc_circ_mat_property:1983}, $U_{\text{left}}$ and $U_{\text{right}}$, both of which are constructed using the DFT matrix:
\begin{equation}\label{eq:decompose_bloc_circ_mat}
M_{\text{bloc-circ}} = U_{\text{left}} \Lambda U_{\text{right}}^\dagger = (F_b \otimes I_p) \cdot \text{diag}(\tilde{B}_0, \tilde{B}_1, \dots, \tilde{B}_{b-1}) \cdot (F_b^\dagger \otimes I_q),
\end{equation}
where $F_b$ is the $b \times b$ DFT matrix.

Each block $\tilde{B}_k$ in the block-diagonal matrix $\text{diag}(\tilde{B}_0, \tilde{B}_1, \dots, \tilde{B}_{b-1})$ is
\begin{equation}
\tilde{B}_k = \sum_{l=0}^{b-1} \alpha^{kl} B_l,
\end{equation}
where $\alpha = e^{-2\pi i / b}$.

We note that the block-circulant matrix defined in Eq.~\ref{eq:def_bloc_circ_mat} is fundamentally different from the block-circulant matrices in Refs.~\cite{bloc_circ_NN:2020,photoelectric_bloc_circ_NN:2024}.
The former consists of a linear number of blocks, with each block containing a quadratic number of independent parameters.
In contrast, the latter (Refs.~\cite{bloc_circ_NN:2020,photoelectric_bloc_circ_NN:2024}) consists of a quadratic number of distinct blocks, each of which is a circulant matrix.

\textbf{Photonic implementation}.
In Eq.~\ref{eq:decompose_bloc_circ_mat}, the middle block-diagonal matrix can be implemented using a quasi-linear number of MZIs, similar to a usual block-diagonal matrix.
The operation $F_b \otimes I_p$ can be realized using a $b \times b$ diffraction unit~\cite{space_efficient_optics:2022}, connected to a $p \times 1$ ($1 \times p$) self-imaging multimode interferometer~\cite{self_imaging_MMI:1995} for each input (output).

\subsection{Performance of NNs with Block-Circulant Weights}
We trained neural networks with block-circulant matrices as weight parameters on the three MNIST-like datasets.
Each network has $5$ bias-free layers with $\left[ 800, 100, 100, 10, 10 \right]$ neurons.
(The input, initially $28 \times 28$ pixels, is flattened and zero-padded to a dimensionality of $800$.)
Each layer, except the last one, is followed by a SiLU activation function.

\begin{figure}[ht]
\centering
    \includegraphics[width=0.9\textwidth]{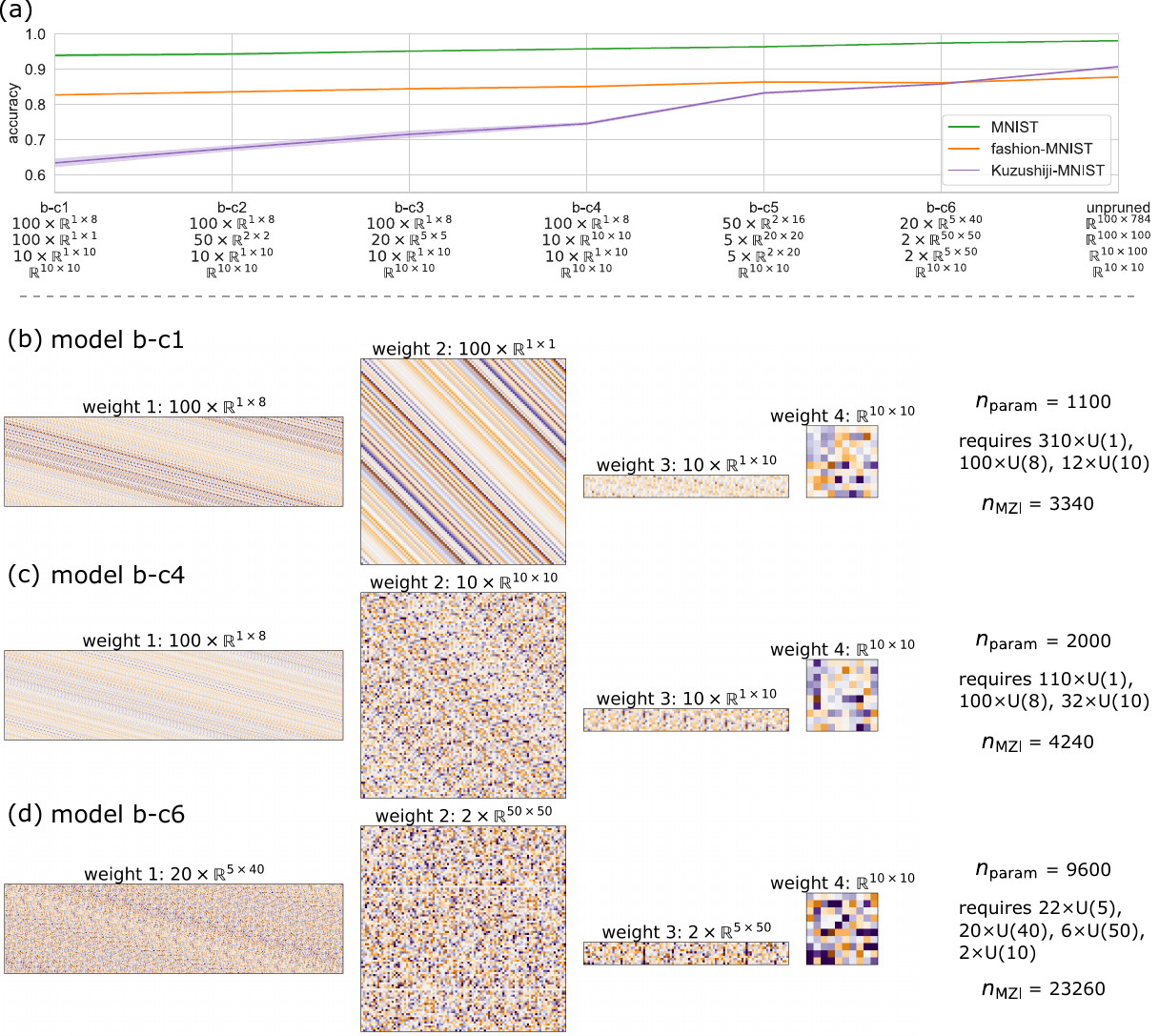}
    \caption{\label{fig:bloc_circ_results}
    (a) The trade-off between inference accuracy and the extent of block-circulant structuring.
    (b-d) Model weights of block-circulant and unpruned models.
    For each model, the number of free parameters, the involved unitary transformations (including both left and right unitary matrices, $U$ and $V^T$, from the singular value decomposition of all block matrices), and the total number of required MZIs are listed on the right.
    }
\end{figure}

Additionally, we implement specific block-circulant layers to perform matrix multiplications using fully vectorized operations, thereby avoiding the trivial construction of the full matrix.
Our approach reshapes the input into blocks, applies cyclic shifts to the blocks of the weight matrix, and performs efficient block-wise multiplication with \texttt{torch.einsum}.
The resulting blocks are then flattened to form the final output.
This implementation significantly improves computational efficiency on GPUs compared to performing matrix operations after reconstructing the full matrix with all cyclically arranged blocks.
We examine six different block-circulant architectures, named from ``b-c1" to ``b-c6."
Fig.~\ref{fig:bloc_circ_results} summarizes the trade-offs between the extent of block-circulant structuring and model accuracy.

These block-circulant models demonstrate slightly higher accuracy compared to their analogous block-diagonal models.
By ``analogous," we refer to the structural similarity and comparable number of independent parameters between the middle block-diagonal matrices in the eigen decomposition of a block-circulant model and a block-diagonal model; for example, the block-circulant model ``b-c1" corresponds to the block-diagonal model ``b-d1" in the main text.
Given that both block-circulant and block-diagonal models have a comparable number of independent parameters, employing block-circulant structured weights is effectively a form of ``pruning."

\section{Algorithms for Constructing the Three Types of Structurally Sparse Matrices}

\begin{algorithm}[H]
\caption{Trivial Sparse Matrix Constructor}\label{alg:TrivialSparseMat}
\begin{algorithmic}[1]
\Procedure{TrivialSparseMat}{$N_{\text{in}}, N_{\text{out}}, \rho$}
\State \Comment{$N_{\text{out}}$: number of output ports, $N_{\text{in}}$: number of input ports, 
$\rho$: the desired density of the matrix}

\State $M \gets \mathbf{0}_{N_{\text{out}} \times N_{\text{in}}}$
\State $flattened\_M \gets$ Flatten $M$ \Comment{Flatten $M$ into a 1D array}

\State $selected\_indices \gets$ random selection of $(\rho \cdot N_{\text{in}} \cdot N_{\text{out}})$ elements from $\{1, 2, \dots, N_{\text{in}} \cdot N_{\text{out}} \}$ without replacement
\State $flattened\_M[selected\_indices] \gets 1$

\State $M \gets$ Reshape $flattened\_M$ to size $(N_{\text{out}}, N_{\text{in}})$ \Comment{Reshape $M$ back to the original dimension}

\State \Return $M$
\EndProcedure
\end{algorithmic}
\end{algorithm}

\begin{algorithm}[H]
\caption{Row Sparse Matrix Constructor}\label{alg:RowSparseMat}
\begin{algorithmic}[1]
\Procedure{RowSparseMat}{$N_{\text{in}}, N_{\text{out}}, \rho, \rho_{\text{row}}$}
\State \Comment{$N_{\text{out}}$: number of output ports, $N_{\text{in}}$: number of input ports, 
$\rho$: the desired density of the matrix,
$\rho_{\text{row}}$: the fraction of rows to be activated}
\State $M \gets \mathbf{0}_{N_{\text{out}} \times N_{\text{in}}}$
\State $indices\_selected\_rows \gets$ random selection of $(\rho_{\text{row}} \cdot N_{\text{out}})$ elements from $\{1, 2, \dots, N_{\text{out}} \}$ without replacement

\If{$\rho$ is `all possible'}
    \State $M[indices\_selected\_rows, :] \gets 1$
\Else
    \State $indices\_entries\_in\_selected\_rows \gets [\ ]$
    \For{$i \in indices\_selected\_rows$}
        \For{$j \gets 1$ \textbf{to} $N_{\text{in}}$}
            \State $indices\_entries\_in\_selected\_rows$.append($(i,j)$)
        \EndFor
    \EndFor
    
    \State $selected\_indices \gets$ random selection of $(\rho \cdot N_{\text{in}} \cdot N_{\text{out}})$ elements from $indices\_entries\_in\_selected\_rows$ without replacement \Comment{Require that $\rho \cdot N_{\text{in}} \cdot N_{\text{out}} \leq$ length$(indices\_entries\_in\_selected\_rows) = \rho_{\text{row}} \cdot N_{\text{in}} \cdot N_{\text{out}}$ to guarantee a valid selection.}

    \For{$(p,q) \in selected\_indices$}
        \State $M_{pq} \gets 1$
    \EndFor
\EndIf

\State \Return $M$
\EndProcedure
\end{algorithmic}
\end{algorithm}

\begin{algorithm}[H]
\caption{2D Local and Sparse Matrix Constructor}\label{alg:2DLSmat}
\begin{algorithmic}[1]
\Procedure{LSmat2D}{$N_{\text{in}}, N_{\text{out}}, \rho, \max(d_{\parallel})$}
\State \Comment{$N_{\text{in}}$: number of input ports, $N_{\text{out}}$: number of output ports,
$\rho$: the desired density of the matrix,
$\max(d_{\parallel})$: the maximum in-plane distance a nonzero coupling can cross}
\State $M \gets \mathbf{0}_{N_{\text{out}} \times N_{\text{in}}}$
\State $indices\_local\_couplings \gets [\ ]$
    \For{$j \gets 1$ \textbf{to} $N_{\text{in}}$}
        \For{$i \gets 1$ \textbf{to} $N_{\text{out}}$}
            \If{$|\mathbf{r}_j - \mathbf{r}_i| \equiv \sqrt{(x_j-x_i)^2+(y_j-y_i)^2} \leq \max(d_{\parallel})$}
                \State $indices\_local\_couplings$.append($(i,j)$)
            \EndIf
        \EndFor
    \EndFor
\If{$\rho$ is `all possible'}
    \State $selected\_indices \gets indices\_local\_couplings$
\Else
    \State $selected\_indices \gets$ random selection of $(\rho \cdot N_{\text{in}} \cdot N_{\text{out}})$ elements from the list $indices\_local\_couplings$ without replacement \Comment{Require that $\rho \cdot N_{\text{in}} \cdot N_{\text{out}} \leq$ length$(indices\_local\_couplings)$ to guarantee a valid selection.}
\EndIf

\For{$(p,q) \in selected\_indices$}
    \State $M_{pq} \gets 1$
\EndFor
\State \Return $M$
\EndProcedure
\end{algorithmic}
\end{algorithm}

\section{Supplementary Dataset}

\begin{table}[hp] 
\centering
\caption{Pruning configurations and performance metrics of the trained conventional ONNs}
\label{tab:fig3_data_conventionalONN}

{\fontsize{9pt}{10pt}\selectfont
\begin{tabular}{c|c|c|c|c|c|C{1.2cm}|C{1.2cm}|C{1.2cm}|c|c}
\toprule
 \multicolumn{11}{c}{conventional ONN, dataset: MNIST} \\
\midrule
 \multicolumn{2}{c|}{threshold} & \multicolumn{3}{c|}{thickness (a.u.)} & \multirow{2}{*}{test accuracy} & \multicolumn{3}{c|}{$\min(l_{\text{cut}})$ for thickness~\cite{details_min_l_cut}} & \multicolumn{2}{c}{row density} \\

 \multicolumn{2}{c|}{$\tau$} & \multicolumn{1}{c}{$t_1$} & \multicolumn{1}{c}{$t_2$} & \multicolumn{1}{c|}{$t_3$} &  & \multicolumn{1}{c}{$1$} & \multicolumn{1}{c}{$2$} & \multicolumn{1}{c|}{$3$} & \multicolumn{1}{c}{$\rho^{(1)}_{\text{row}}$} & \multicolumn{1}{c}{$\rho^{(2)}_{\text{row}}$} \\

\hline
\multicolumn{2}{c|}{$0.05$} & $14.51 \pm 0.38$ & $19.21 \pm 0.85$ & $0.87 \pm 0.24$ & $97.67 \pm 0.15\%$ & $5.66$ & $4.24$ & $4.24$ & $1.00$ & $1.00$ \\
\multicolumn{2}{c|}{$0.075$} & $13.96 \pm 0.16$ & $16.56 \pm 1.42$ & $0.73 \pm 0.28$ & $95.58 \pm 2.97\%$ & $6.32$ & $4.24$ & $4.24$ & $1.00$ & $1.00$ \\
\multicolumn{2}{c|}{$0.1$} & $13.63 \pm 0.10$ & $14.20 \pm 1.56$ & $0.50 \pm 0.12$ & $93.41 \pm 4.10\%$ & $7.07$ & $4.24$ & $4.24$ & $1.00$ & $1.00$ \\
\multicolumn{2}{c|}{$0.15$} & $12.80 \pm 0.25$ & $10.97 \pm 2.20$ & $0.42 \pm 0.15$ & $79.87 \pm 6.72\%$ & $7.07$ & $4.24$ & $4.24$ & $1.00$ & $1.00$ \\
\multicolumn{2}{c|}{$0.2$} & $11.92 \pm 0.34$ & $8.72 \pm 1.99$ & $0.30 \pm 0.08$ & $59.11 \pm 13.03\%$ & $7.07$ & $4.24$ & $7.81$ & $1.00$ & $1.00$ \\

\midrule
 \multicolumn{11}{c}{} \\
 \multicolumn{11}{c}{conventional ONN, dataset: fashion-MNIST} \\
\midrule
 \multicolumn{2}{c|}{threshold} & \multicolumn{3}{c|}{thickness (a.u.)} & \multirow{2}{*}{test accuracy} & \multicolumn{3}{c|}{$\min(l_{\text{cut}})$ for thickness} & \multicolumn{2}{c}{row density} \\

 \multicolumn{2}{c|}{$\tau$} & \multicolumn{1}{c}{$t_1$} & \multicolumn{1}{c}{$t_2$} & \multicolumn{1}{c|}{$t_3$} &  & \multicolumn{1}{c}{$1$} & \multicolumn{1}{c}{$2$} & \multicolumn{1}{c|}{$3$} & \multicolumn{1}{c}{$\rho^{(1)}_{\text{row}}$} & \multicolumn{1}{c}{$\rho^{(2)}_{\text{row}}$} \\

\hline
\multicolumn{2}{c|}{$0.05$} & $65.14 \pm 1.34$ & $18.56 \pm 1.22$ & $1.18 \pm 0.48$ & $88.51 \pm 0.28\%$ & $1.41$ & $4.24$ & $4.24$ & $1.00$ & $1.00$ \\
\multicolumn{2}{c|}{$0.075$} & $62.84 \pm 1.43$ & $16.20 \pm 1.50$ & $0.80 \pm 0.24$ & $88.08 \pm 0.36\%$ & $1.41$ & $4.24$ & $4.24$ & $1.00$ & $1.00$ \\
\multicolumn{2}{c|}{$0.1$} & $60.72 \pm 2.16$ & $13.94 \pm 1.77$ & $0.60 \pm 0.15$ & $85.82 \pm 1.69\%$ & $1.41$ & $4.24$ & $4.24$ & $1.00$ & $1.00$ \\
\multicolumn{2}{c|}{$0.15$} & $56.48 \pm 2.19$ & $10.31 \pm 1.91$ & $0.40 \pm 0.05$ & $73.84 \pm 5.89\%$ & $1.41$ & $4.24$ & $4.24$ & $1.00$ & $1.00$ \\
\multicolumn{2}{c|}{$0.2$} & $53.12 \pm 2.88$ & $8.28 \pm 1.38$ & $0.35 \pm 0.05$ & $47.31 \pm 7.97\%$ & $1.41$ & $4.24$ & $4.24$ & $1.00$ & $1.00$ \\

\midrule
 \multicolumn{11}{c}{} \\
 \multicolumn{11}{c}{conventional ONN, dataset: Kuzushiji-MNIST} \\
\midrule
 \multicolumn{2}{c|}{threshold} & \multicolumn{3}{c|}{thickness (a.u.)} & \multirow{2}{*}{test accuracy} & \multicolumn{3}{c|}{$\min(l_{\text{cut}})$ for thickness} & \multicolumn{2}{c}{row density} \\

 \multicolumn{2}{c|}{$\tau$} & \multicolumn{1}{c}{$t_1$} & \multicolumn{1}{c}{$t_2$} & \multicolumn{1}{c|}{$t_3$} &  & \multicolumn{1}{c}{$1$} & \multicolumn{1}{c}{$2$} & \multicolumn{1}{c|}{$3$} & \multicolumn{1}{c}{$\rho^{(1)}_{\text{row}}$} & \multicolumn{1}{c}{$\rho^{(2)}_{\text{row}}$} \\

\hline
\multicolumn{2}{c|}{$0.05$} & $66.73 \pm 1.27$ & $17.38 \pm 1.15$ & $0.74 \pm 0.24$ & $87.28 \pm 0.43\%$ & $1.41$ & $4.24$ & $4.24$ & $1.00$ & $1.00$ \\
\multicolumn{2}{c|}{$0.075$} & $63.73 \pm 1.64$ & $14.76 \pm 1.87$ & $0.49 \pm 0.11$ & $82.18 \pm 5.05\%$ & $1.41$ & $4.24$ & $4.24$ & $1.00$ & $1.00$ \\
\multicolumn{2}{c|}{$0.1$} & $60.46 \pm 2.09$ & $13.17 \pm 1.54$ & $0.41 \pm 0.05$ & $76.46 \pm 7.51\%$ & $1.41$ & $4.24$ & $4.24$ & $1.00$ & $1.00$ \\
\multicolumn{2}{c|}{$0.15$} & $54.62 \pm 3.22$ & $9.46 \pm 1.75$ & $0.33 \pm 0.04$ & $61.22 \pm 7.01\%$ & $1.41$ & $4.24$ & $12.04$ & $1.00$ & $1.00$ \\
\multicolumn{2}{c|}{$0.2$} & $49.50 \pm 2.65$ & $7.19 \pm 1.41$ & $0.26 \pm 0.03$ & $51.23 \pm 5.60\%$ & $1.41$ & $4.24$ & $4.24$ & $1.00$ & $1.00$ \\

 \bottomrule
\end{tabular}
}
\end{table}

\begin{table}[hp]
\centering
\caption{Pruning configurations and performance metrics of the trained row sparse ONNs}
\label{tab:fig3_data_sparseONN}

{\fontsize{9pt}{10pt}\selectfont
\begin{tabular}{c|c|c|c|c|c|C{1.2cm}|C{1.2cm}|C{1.2cm}|c|c}
\toprule
 \multicolumn{11}{c}{row sparse ONN, dataset: MNIST} \\
\midrule
 \multicolumn{2}{c|}{threshold} & \multicolumn{3}{c|}{thickness (a.u.)} & \multirow{2}{*}{test accuracy} & \multicolumn{3}{c|}{$\min(l_{\text{cut}})$ for thickness~\cite{details_min_l_cut}} & \multicolumn{2}{c}{row density} \\

 \multicolumn{1}{c}{$\tau$} & \multicolumn{1}{c|}{$\tau_{\text{neuron}}$} & \multicolumn{1}{c}{$t_1$} & \multicolumn{1}{c}{$t_2$} & \multicolumn{1}{c|}{$t_3$} &  & \multicolumn{1}{c}{$1$} & \multicolumn{1}{c}{$2$} & \multicolumn{1}{c|}{$3$} & \multicolumn{1}{c}{$\rho^{(1)}_{\text{row}}$} & \multicolumn{1}{c}{$\rho^{(2)}_{\text{row}}$} \\

\hline
$0.05$ & $0.05$ & $11.82 \pm 0.48$ & $12.96 \pm 2.07$ & $0.86 \pm 0.25$ & $96.63 \pm 0.39\%$ & $5.66$ & $4.24$ & $4.24$ & $0.81 \pm 0.03$ & $0.70 \pm 0.10$ \\
$0.075$ & $0.075$ & $10.05 \pm 0.43$ & $8.01 \pm 1.83$ & $0.69 \pm 0.30$ & $91.39 \pm 5.54\%$ & $6.32$ & $4.24$ & $4.24$ & $0.72 \pm 0.03$ & $0.51 \pm 0.11$ \\
$0.1$ & $0.1$ & $8.82 \pm 0.57$ & $4.58 \pm 1.00$ & $0.45 \pm 0.14$ & $86.64 \pm 9.79\%$ & $6.32$ & $4.24$ & $4.24$ & $0.64 \pm 0.04$ & $0.38 \pm 0.07$ \\
$0.125$ & $0.1$ & $7.96 \pm 0.83$ & $4.01 \pm 1.02$ & $0.45 \pm 0.14$ & $86.74 \pm 9.50\%$ & $6.32$ & $4.24$ & $4.24$ & $0.58 \pm 0.06$ & $0.31 \pm 0.06$ \\
$0.15$ & $0.15$ & $6.84 \pm 0.81$ & $2.39 \pm 0.48$ & $0.37 \pm 0.18$ & $70.01 \pm 5.21\%$ & $7.07$ & $4.24$ & $4.24$ & $0.53 \pm 0.06$ & $0.26 \pm 0.05$ \\

\midrule
 \multicolumn{11}{c}{} \\
 \multicolumn{11}{c}{row sparse ONN, dataset: fashion-MNIST} \\
\midrule
 \multicolumn{2}{c|}{threshold} & \multicolumn{3}{c|}{thickness (a.u.)} & \multirow{2}{*}{test accuracy} & \multicolumn{3}{c|}{$\min(l_{\text{cut}})$ for thickness} & \multicolumn{2}{c}{row density} \\

 \multicolumn{1}{c}{$\tau$} & \multicolumn{1}{c|}{$\tau_{\text{neuron}}$} & \multicolumn{1}{c}{$t_1$} & \multicolumn{1}{c}{$t_2$} & \multicolumn{1}{c|}{$t_3$} &  & \multicolumn{1}{c}{$1$} & \multicolumn{1}{c}{$2$} & \multicolumn{1}{c|}{$3$} & \multicolumn{1}{c}{$\rho^{(1)}_{\text{row}}$} & \multicolumn{1}{c}{$\rho^{(2)}_{\text{row}}$} \\

\hline
$0.05$ & $0.05$ & $42.43 \pm 3.62$ & $14.08 \pm 2.14$ & $1.15 \pm 0.50$ & $86.04 \pm 0.65\%$ & $1.41$ & $4.24$ & $4.24$ & $0.64 \pm 0.06$ & $0.79 \pm 0.06$ \\
$0.075$ & $0.075$ & $34.38 \pm 2.80$ & $8.81 \pm 2.05$ & $0.77 \pm 0.26$ & $82.84 \pm 1.58\%$ & $1.41$ & $4.24$ & $4.24$ & $0.53 \pm 0.05$ & $0.61 \pm 0.09$ \\
$0.1$ & $0.1$ & $27.58 \pm 2.42$ & $5.07 \pm 1.30$ & $0.58 \pm 0.17$ & $74.35 \pm 5.07\%$ & $1.41$ & $4.24$ & $4.24$ & $0.43 \pm 0.04$ & $0.45 \pm 0.07$ \\
$0.125$ & $0.1$ & $23.51 \pm 2.14$ & $3.30 \pm 0.79$ & $0.47 \pm 0.09$ & $68.51 \pm 4.75\%$ & $1.41$ & $4.24$ & $4.24$ & $0.37 \pm 0.03$ & $0.31 \pm 0.05$ \\
$0.15$ & $0.15$ & $20.95 \pm 1.77$ & $1.56 \pm 0.54$ & $0.30 \pm 0.02$ & $45.63 \pm 7.41\%$ & $1.41$ & $4.24$ & $17.46$ & $0.35 \pm 0.03$ & $0.20 \pm 0.02$ \\

\midrule
 \multicolumn{11}{c}{} \\
 \multicolumn{11}{c}{row sparse ONN, dataset: Kuzushiji-MNIST} \\
\midrule
 \multicolumn{2}{c|}{threshold} & \multicolumn{3}{c|}{thickness (a.u.)} & \multirow{2}{*}{test accuracy} & \multicolumn{3}{c|}{$\min(l_{\text{cut}})$ for thickness} & \multicolumn{2}{c}{row density} \\

 \multicolumn{1}{c}{$\tau$} & \multicolumn{1}{c|}{$\tau_{\text{neuron}}$} & \multicolumn{1}{c}{$t_1$} & \multicolumn{1}{c}{$t_2$} & \multicolumn{1}{c|}{$t_3$} &  & \multicolumn{1}{c}{$1$} & \multicolumn{1}{c}{$2$} & \multicolumn{1}{c|}{$3$} & \multicolumn{1}{c}{$\rho^{(1)}_{\text{row}}$} & \multicolumn{1}{c}{$\rho^{(2)}_{\text{row}}$} \\

\hline
$0.05$ & $0.05$ & $60.99 \pm 2.37$ & $10.43 \pm 0.83$ & $0.66 \pm 0.26$ & $85.63 \pm 0.54\%$ & $1.41$ & $4.24$ & $4.24$ & $0.91 \pm 0.03$ & $0.65 \pm 0.08$ \\
$0.075$ & $0.075$ & $56.04 \pm 2.37$ & $6.95 \pm 0.92$ & $0.45 \pm 0.13$ & $77.46 \pm 6.52\%$ & $1.41$ & $4.24$ & $4.24$ & $0.87 \pm 0.03$ & $0.54 \pm 0.09$ \\
$0.1$ & $0.1$ & $49.50 \pm 2.29$ & $5.27 \pm 0.98$ & $0.37 \pm 0.06$ & $69.31 \pm 8.19\%$ & $1.41$ & $4.24$ & $4.24$ & $0.82 \pm 0.03$ & $0.47 \pm 0.08$ \\
$0.125$ & $0.1$ & $47.11 \pm 2.32$ & $4.71 \pm 0.82$ & $0.37 \pm 0.06$ & $67.46 \pm 8.48\%$ & $1.41$ & $4.24$ & $4.24$ & $0.78 \pm 0.02$ & $0.41 \pm 0.06$ \\
$0.15$ & $0.15$ & $40.31 \pm 3.02$ & $3.43 \pm 0.65$ & $0.31 \pm 0.03$ & $53.24 \pm 6.02\%$ & $1.41$ & $4.24$ & $24.04$ & $0.72 \pm 0.03$ & $0.37 \pm 0.05$ \\

 \bottomrule
\end{tabular}
}
\end{table}

\begin{table}[hp]
\centering
\caption{Pruning configurations and performance metrics of the trained LSONNs}
\label{tab:fig3_data_LSONN}

{\fontsize{9pt}{10pt}\selectfont
\begin{tabular}{c|c|c|c|c|c|C{1.2cm}|C{1.2cm}|C{1.2cm}|c|c}
\toprule
 \multicolumn{11}{c}{LSONN, dataset: MNIST} \\
\midrule
 \multicolumn{2}{c|}{threshold} & \multicolumn{3}{c|}{thickness (a.u.)} & \multirow{2}{*}{test accuracy} & \multicolumn{3}{c|}{$\min(l_{\text{cut}})$ for thickness~\cite{details_min_l_cut}} & \multicolumn{2}{c}{row density} \\

 \multicolumn{2}{c|}{$\tau$} & \multicolumn{1}{c}{$t_1$} & \multicolumn{1}{c}{$t_2$} & \multicolumn{1}{c|}{$t_3$} &  & \multicolumn{1}{c}{$1$} & \multicolumn{1}{c}{$2$} & \multicolumn{1}{c|}{$3$} & \multicolumn{1}{c}{$\rho^{(1)}_{\text{row}}$} & \multicolumn{1}{c}{$\rho^{(2)}_{\text{row}}$} \\

\hline
\multicolumn{2}{c|}{$0.01$} & $1.15 \pm 0.08$ & $0.40 \pm 0.03$ & $0.26 \pm 0.03$ & $95.86 \pm 0.22\%$ & $23.20$ & $28.00$ & $28.00$ & $0.31 \pm 0.02$ & $0.12 \pm 0.01$ \\
\multicolumn{2}{c|}{$0.02$} & $1.12 \pm 0.08$ & $0.40 \pm 0.03$ & $0.21 \pm 0.06$ & $95.79 \pm 0.22\%$ & $24.04$ & $28.00$ & $27.59$ & $0.31 \pm 0.02$ & $0.12 \pm 0.01$ \\
\multicolumn{2}{c|}{$0.05$} & $1.12 \pm 0.08$ & $0.39 \pm 0.03$ & $0.18 \pm 0.05$ & $95.37 \pm 0.52\%$ & $28.00$ & $28.00$ & $27.59$ & $0.31 \pm 0.02$ & $0.12 \pm 0.01$ \\
\multicolumn{2}{c|}{$0.1$} & $1.11 \pm 0.07$ & $0.39 \pm 0.02$ & $0.16 \pm 0.04$ & $93.35 \pm 0.61\%$ & $28.00$ & $28.00$ & $28.00$ & $0.31 \pm 0.02$ & $0.12 \pm 0.01$ \\
\multicolumn{2}{c|}{$0.2$} & $1.03 \pm 0.05$ & $0.34 \pm 0.01$ & $0.14 \pm 0.02$ & $71.09 \pm 7.33\%$ & $28.00$ & $28.00$ & $27.59$ & $0.31 \pm 0.02$ & $0.11 \pm 0.01$ \\

\midrule
 \multicolumn{11}{c}{} \\
 \multicolumn{11}{c}{LSONN, dataset: fashion-MNIST} \\
\midrule
 \multicolumn{2}{c|}{threshold} & \multicolumn{3}{c|}{thickness (a.u.)} & \multirow{2}{*}{test accuracy} & \multicolumn{3}{c|}{$\min(l_{\text{cut}})$ for thickness} & \multicolumn{2}{c}{row density} \\

 \multicolumn{2}{c|}{$\tau$} & \multicolumn{1}{c}{$t_1$} & \multicolumn{1}{c}{$t_2$} & \multicolumn{1}{c|}{$t_3$} &  & \multicolumn{1}{c}{$1$} & \multicolumn{1}{c}{$2$} & \multicolumn{1}{c|}{$3$} & \multicolumn{1}{c}{$\rho^{(1)}_{\text{row}}$} & \multicolumn{1}{c}{$\rho^{(2)}_{\text{row}}$} \\

\hline
\multicolumn{2}{c|}{$0.01$} & $1.12 \pm 0.07$ & $0.40 \pm 0.03$ & $0.34 \pm 0.02$ & $84.95 \pm 0.24\%$ & $14.14$ & $27.93$ & $28.00$ & $0.25 \pm 0.02$ & $0.12 \pm 0.01$ \\
\multicolumn{2}{c|}{$0.02$} & $1.07 \pm 0.08$ & $0.39 \pm 0.03$ & $0.29 \pm 0.05$ & $84.88 \pm 0.19\%$ & $13.21$ & $27.93$ & $28.00$ & $0.25 \pm 0.02$ & $0.12 \pm 0.01$ \\
\multicolumn{2}{c|}{$0.05$} & $1.00 \pm 0.06$ & $0.38 \pm 0.02$ & $0.22 \pm 0.04$ & $84.30 \pm 0.26\%$ & $15.56$ & $27.93$ & $28.02$ & $0.25 \pm 0.02$ & $0.12 \pm 0.01$ \\
\multicolumn{2}{c|}{$0.1$} & $0.95 \pm 0.07$ & $0.37 \pm 0.03$ & $0.14 \pm 0.01$ & $78.41 \pm 2.24\%$ & $15.56$ & $27.93$ & $27.59$ & $0.25 \pm 0.02$ & $0.12 \pm 0.01$ \\
\multicolumn{2}{c|}{$0.2$} & $0.81 \pm 0.06$ & $0.35 \pm 0.02$ & $0.12 \pm 0.01$ & $50.66 \pm 5.46\%$ & $16.38$ & $28.00$ & $27.59$ & $0.24 \pm 0.02$ & $0.11 \pm 0.01$ \\

\midrule
 \multicolumn{11}{c}{} \\
 \multicolumn{11}{c}{LSONN, dataset: Kuzushiji-MNIST} \\
\midrule
 \multicolumn{2}{c|}{threshold} & \multicolumn{3}{c|}{thickness (a.u.)} & \multirow{2}{*}{test accuracy} & \multicolumn{3}{c|}{$\min(l_{\text{cut}})$ for thickness} & \multicolumn{2}{c}{row density} \\

 \multicolumn{2}{c|}{$\tau$} & \multicolumn{1}{c}{$t_1$} & \multicolumn{1}{c}{$t_2$} & \multicolumn{1}{c|}{$t_3$} &  & \multicolumn{1}{c}{$1$} & \multicolumn{1}{c}{$2$} & \multicolumn{1}{c|}{$3$} & \multicolumn{1}{c}{$\rho^{(1)}_{\text{row}}$} & \multicolumn{1}{c}{$\rho^{(2)}_{\text{row}}$} \\

\hline
\multicolumn{2}{c|}{$0.01$} & $2.00 \pm 0.12$ & $0.44 \pm 0.03$ & $0.34 \pm 0.02$ & $78.63 \pm 0.89\%$ & $12.73$ & $19.80$ & $28.00$ & $0.39 \pm 0.02$ & $0.13 \pm 0.01$ \\
\multicolumn{2}{c|}{$0.02$} & $1.77 \pm 0.13$ & $0.43 \pm 0.04$ & $0.21 \pm 0.03$ & $78.60 \pm 0.85\%$ & $14.14$ & $28.00$ & $28.02$ & $0.39 \pm 0.02$ & $0.13 \pm 0.01$ \\
\multicolumn{2}{c|}{$0.05$} & $1.65 \pm 0.12$ & $0.43 \pm 0.04$ & $0.15 \pm 0.04$ & $77.79 \pm 0.83\%$ & $18.38$ & $28.00$ & $27.59$ & $0.39 \pm 0.02$ & $0.13 \pm 0.01$ \\
\multicolumn{2}{c|}{$0.1$} & $1.45 \pm 0.08$ & $0.40 \pm 0.02$ & $0.13 \pm 0.02$ & $72.32 \pm 1.70\%$ & $15.56$ & $28.00$ & $27.59$ & $0.39 \pm 0.02$ & $0.13 \pm 0.01$ \\
\multicolumn{2}{c|}{$0.2$} & $1.20 \pm 0.06$ & $0.38 \pm 0.04$ & $0.11 \pm 0.00$ & $40.02 \pm 3.36\%$ & $28.00$ & $28.00$ & $27.34$ & $0.39 \pm 0.02$ & $0.13 \pm 0.01$ \\

 \bottomrule
\end{tabular}
}
\end{table}

\begin{table}[hp]
\centering
\caption{Comparison of conventional and local sparse ONNs}
\label{tab:fig3_data_thickness_reduction}

{\fontsize{9pt}{10pt}\selectfont
\begin{tabular}{C{3.0cm}|C{3.0cm}|C{2.0cm}|C{2.0cm}|C{2.0cm}|c}
\toprule
 \multicolumn{6}{c}{dataset: MNIST} \\
\midrule
 \multicolumn{2}{c|}{pruning threshold} & \multicolumn{3}{c|}{thickness reduction factor} & accuracy \\

 conventional ONN & local sparse ONN & \multicolumn{1}{c}{$t_1/t^{\text{ls}}_1$} & \multicolumn{1}{c}{$t_2/t^{\text{ls}}_2$} & \multicolumn{1}{c|}{$t_3/t^{\text{ls}}_3$} & degradation \\

\hline
$0.05$ & $0.01$ & $12.6 \pm 1.0$ & $47.8 \pm 4.1$ & $3.4 \pm 1.0$ & $1.81 \pm 0.27 \%$ \\
$0.075$ & $0.02$ & $12.5 \pm 0.8$ & $41.2 \pm 4.7$ & $3.5 \pm 1.7$ & $-0.22 \pm 2.98 \%$ \\
$0.1$ & $0.05$ & $12.2 \pm 0.8$ & $36.2 \pm 4.6$ & $2.7 \pm 1.0$ & $-1.96 \pm 4.13 \%$ \\
$0.15$ & $0.1$ & $11.5 \pm 0.7$ & $28.4 \pm 5.9$ & $2.5 \pm 1.1$ & $-13.48 \pm 6.74 \%$ \\
$0.2$ & $0.2$ & $11.6 \pm 0.6$ & $25.4 \pm 5.9$ & $2.2 \pm 0.7$ & $-11.97 \pm 14.95 \%$ \\

\midrule
 \multicolumn{6}{c}{} \\
 \multicolumn{6}{c}{dataset: fashion-MNIST} \\
\midrule
 \multicolumn{2}{c|}{pruning threshold} & \multicolumn{3}{c|}{thickness reduction factor} & accuracy \\

 conventional ONN & local sparse ONN & \multicolumn{1}{c}{$t_1/t^{\text{ls}}_1$} & \multicolumn{1}{c}{$t_2/t^{\text{ls}}_2$} & \multicolumn{1}{c|}{$t_3/t^{\text{ls}}_3$} & degradation \\

\hline
$0.05$ & $0.01$ & $58.1 \pm 4.0$ & $46.3 \pm 4.6$ & $3.5 \pm 1.4$ & $3.55 \pm 0.36 \%$ \\
$0.075$ & $0.02$ & $58.8 \pm 4.9$ & $41.1 \pm 4.6$ & $2.8 \pm 1.0$ & $3.19 \pm 0.41 \%$ \\
$0.1$ & $0.05$ & $61.0 \pm 4.4$ & $36.4 \pm 5.1$ & $2.7 \pm 0.9$ & $1.52 \pm 1.71 \%$ \\
$0.15$ & $0.1$ & $59.7 \pm 4.8$ & $28.0 \pm 5.6$ & $2.9 \pm 0.5$ & $-4.57 \pm 6.30 \%$ \\
$0.2$ & $0.2$ & $65.5 \pm 6.2$ & $23.3 \pm 4.2$ & $2.9 \pm 0.5$ & $-3.35 \pm 9.66 \%$ \\

\midrule
 \multicolumn{6}{c}{} \\
 \multicolumn{6}{c}{dataset: Kuzushiji-MNIST} \\
\midrule
 \multicolumn{2}{c|}{pruning threshold} & \multicolumn{3}{c|}{thickness reduction factor} & accuracy \\

 conventional ONN & local sparse ONN & \multicolumn{1}{c}{$t_1/t^{\text{ls}}_1$} & \multicolumn{1}{c}{$t_2/t^{\text{ls}}_2$} & \multicolumn{1}{c|}{$t_3/t^{\text{ls}}_3$} & degradation \\

\hline
$0.05$ & $0.01$ & $33.4 \pm 2.1$ & $39.2 \pm 4.0$ & $2.2 \pm 0.7$ & $8.65 \pm 0.99 \%$ \\
$0.075$ & $0.02$ & $35.9 \pm 2.8$ & $34.3 \pm 5.2$ & $2.3 \pm 0.6$ & $3.58 \pm 5.12 \%$ \\
$0.1$ & $0.05$ & $36.6 \pm 2.9$ & $30.8 \pm 4.4$ & $2.7 \pm 0.7$ & $-1.33 \pm 7.56 \%$ \\
$0.15$ & $0.1$ & $37.6 \pm 3.0$ & $23.4 \pm 4.6$ & $2.6 \pm 0.5$ & $-11.10 \pm 7.22 \%$ \\
$0.2$ & $0.2$ & $41.3 \pm 3.1$ & $18.7 \pm 4.1$ & $2.4 \pm 0.3$ & $11.21 \pm 6.53 \%$ \\

 \bottomrule
\end{tabular}
}
\end{table}

\begin{figure}[ht]
\centering
    \includegraphics[width=1\textwidth]{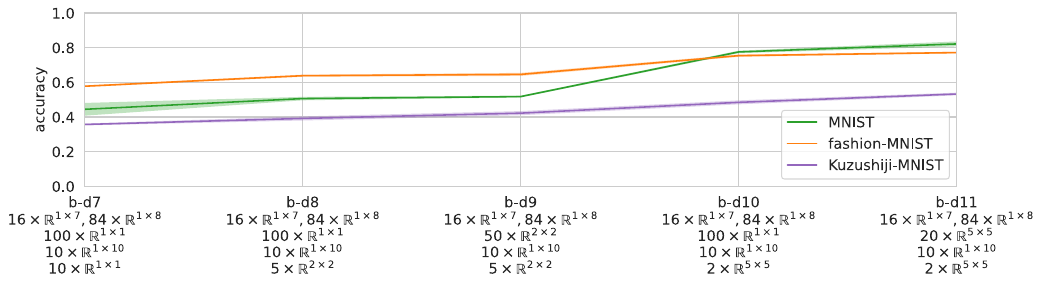}
    \caption{\label{fig:extended_data__bloc_diag_results}
    The trade-off between inference accuracy and the extent of block-diagonalization for models with the last weight matrix block-diagonalized.
    Shaded regions represent one standard deviation of accuracy across eight different random seeds.
    }
\end{figure}

\begin{table}[hp]
  \centering
  \renewcommand{\arraystretch}{1.5}
  \begin{tabular}{c|c|c|c|c|c}
    \hline
    input dim & operator & $t$ & $c_{\text{out}}$ & $n$ & $s$ \\ \hline\hline
    $32^2 \times 3$ & conv2d & - & 32 & 1 & 1 \\
    $32^2 \times 32$ & bottleneck & 1 & 16 & 1 & 1 \\
    $32^2 \times 16$ & bottleneck & 6 & 24 & 2 & 1 \\
    $32^2 \times 24$ & bottleneck & 6 & 32 & 3 & 1 \\
    $32^2 \times 32$ & bottleneck & 6 & 64 & 4 & 2 \\
    $16^2 \times 64$ & bottleneck & 6 & 96 & 3 & 1 \\
    $16^2 \times 96$ & bottleneck & 6 & 160 & 3 & 2 \\
    $8^2 \times 160$ & bottleneck & 6 & 320 & 1 & 1 \\
    $8^2 \times 320$ & conv2d 1x1 & - & 1280 & 1 & 1 \\
    $8^2 \times 1280$ & avgpool 8x8 & - & - & 1 & - \\ \hline
  \end{tabular}
  \caption{Architecture of the modified MobileNetV2, with and without pruning.
  The last fully connected layer(s) is not shown.
  Parameters for \textit{bottleneck layers}: the expansion factor $t$, output channels $c$, and stride $s$.
  The bottleneck layer transforms an input of dimensions $h \times w \times c_{\text{in}}$ to $(h/s) \times (w/s) \times c_{\text{out}}$~\cite{MobileNetV2:2018}.
  $n$ represents how many times the bottleneck layer repeats.
  When repeating a bottleneck layer with downsampling (i.e., $s > 1$), the downsampling only applies the first time that bottleneck layer is used.
  }
  \label{tab:mobilenetv2_architecture}
\end{table}

\clearpage
\bibliographystyle{unsrt}